\documentclass[preprint,12pt, a4paper, dvipsnames]{elsarticle}

\usepackage{amssymb}

\usepackage{amsthm}
\usepackage{mathtools}

\usepackage{lineno}
\usepackage{listings}

\usepackage{float}

\usepackage{url}

\usepackage{xcolor}

\usepackage{caption}
\usepackage{subcaption}

\definecolor{codegreen}{rgb}{0.25,0.93,0.31}
\definecolor{codegray}{rgb}{0.5,0.5,0.5}
\definecolor{codepurple}{rgb}{0.58,0,0.82}
\definecolor{backcolour}{rgb}{0.95,0.95,0.92}

\restylefloat{table}

\newcommand{\ie}{{\emph{i.e.\/}}}
\newcommand{\ket}[1]{\ensuremath{|#1\rangle}}
\newcommand{\bra}[1]{\ensuremath{\langle#1|}}
\newcommand{\ketbra}[2]{\ensuremath{\ket{#1}\bra{#2}}}
\newcommand{\proj}[1]{\ensuremath{\ketbra{#1}{#1}}}
\newcommand{\braket}[2]{\ensuremath{\langle{#1}|{#2}\rangle}}

\newcommand{\1}{{\rm 1\hspace{-0.9mm}l}}
\newcommand{\Id}{{\rm 1\hspace{-0.9mm}l}}

\newcommand{\DD}{\mathcal{D}}

\newcommand{\PP}{\mathcal{P}}
\newcommand{\QQ}{\mathcal{Q}}

\newcommand{\UU}{\mathcal{U}}

\newcommand{\diaguni}{\ensuremath{\mathcal{DU}}}
\newcommand{\diag}{\mathrm{diag}}
\newcommand{\tr}{\mathrm{tr}}
\newcommand{\textapprox}{\raisebox{0.5ex}{\texttildelow}}
\journal{SoftwareX}

\usepackage{amsmath}
\newtheorem{theorem}{Theorem}
\newtheorem{proposition}{Proposition}

\newtheorem{lemma}{Lemma}
\theoremstyle{definition}
\newtheorem{scheme}{Scheme}

\begin{document}

\begin{frontmatter}

\title{PyQBench: a Python library for benchmarking gate-based quantum computers}

\author{Konrad Jałowiecki\corref{cor1}}
\ead{dexter2206@gmail.com}
\cortext[cor1]{Corresponding author}

\author{Paulina Lewandowska}
\author{\L ukasz Pawela}

\address{Institute of Theoretical and Applied Informatics, Polish Academy
	of Sciences, Ba{\l}tycka~5, 44-100 Gliwice, Poland}

\begin{abstract}

We introduce PyQBench, an innovative open-source framework for benchmarking
gate-based quantum computers. PyQBench can benchmark NISQ devices by verifying their capability of
discriminating between two von Neumann measurements. PyQBench offers a simplified, ready-to-use,
command line interface (CLI) for running benchmarks using a predefined parametrized Fourier
family of measurements. For more advanced scenarios, PyQBench offers a way of employing user-defined
measurements instead of predefined ones.

\end{abstract}


\end{frontmatter}

\section*{Current code version}
\label{}

\begin{table}[H]
\begin{tabular}{|l|p{5.5cm}|p{7.5cm}|}
\hline
C1 & Current code version & 0.1.1 \\
\hline
C2 & Permanent link to code/repository used for this code version & \url{https://github.com/iitis/PyQBench} \\
\hline
C3 & Code Ocean compute capsule & \texttt{https://codeocean.com/capsule/ 89088992-9a27-4712-8525-} \texttt{d92a9b23060f/tree}\\
\hline
C4 & Legal Code License & Apache License 2.0\\
\hline
C5 & Code versioning system used & git \\
\hline
C6 & Software code languages, tools, and services used & Python, Qiskit, AWS Braket \\
\hline
C7 & Compilation requirements, operating environments \& dependencies &
\texttt{Python >= 3.8}\newline
\texttt{numpy \textapprox= 1.22.0}\newline
\texttt{scipy \textapprox= 1.7.0}\newline
\texttt{pandas \textapprox= 1.5.0}\newline
\texttt{amazon-braket-sdk >= 1.11.1}\newline
\texttt{pydantic \textapprox= 1.9.1}\newline
\texttt{qiskit \textapprox= 0.37.2}\newline
\texttt{mthree \textapprox= 1.1.0}\newline
\texttt{tqdm \textapprox= 4.64.1}\newline
\texttt{pyyaml \textapprox= 6.0}\newline
\texttt{qiskit-braket-provider \textapprox= 0.0.3}\\
\hline
C8 & If available Link to developer documentation/manual &
\url{https://pyqbench.readthedocs.io/en/latest/}\\
\hline
C9 & Support email for questions & \url{dexter2206@gmail.com}\\
\hline
\end{tabular}
\caption{Code metadata}
\label{}
\end{table}

\section{Motivation and significance}

Noisy Intermediate-Scale Quantum (NISQ)~\cite{preskill} devices are storming the market,
with a wide selection of devices based on different architectures and accompanying software
solutions. Among hardware providers offering public access to their gate--based devices, one could
mention Rigetti \cite{rigetti}, IBM \cite{ibmq}, Oxford Quantum Group \cite{oxforf}, IonQ \cite{ionq} or Xanadu \cite{xanadu}. Other vendors offer devices operating in
different paradigms. Notably, one could mention D-Wave \cite{dwave} and their quantum
annealers, or QuEra devices \cite{quera} based on neural atoms.  Most vendors provide their own software stack and
application programming interface for accessing their devices. To name a few, Rigetti's computers
are available through their Forest SDK \cite{sdk} and PyQuil library \cite{pyquil} and IBM Q \cite{ibmq} computers can be accessed
through Qiskit \cite{qiskit} or IBM Quantum Experience web interface \cite{ibmqplatform}. Some cloud services, like Amazon Braket \cite{amazon}, offer access to several quantum devices under a unified API. On top of that, several
libraries and frameworks can integrate with multiple hardware vendors. Examples of such frameworks
include IBM Q's Qiskit or Zapata Computing's Orquestra \cite{zapata}.

It is well known that NISQ devices have their limitations \cite{preskillnew}. The question is to what extent those
devices can perform meaningful computations? To answer this question, one has to devise a
methodology for benchmarking them. For gate--based computers, on which this paper focuses, there
already exist several approaches.
One could mention randomized benchmarking \cite{liu2022sampling, knill2008randomized, wallman2014randomized, helsen2022general,
	cornelissen2021scalable}, benchmarks based on the quantum volume \cite{cross2019validating, moll2018quantum, pelofske2022volume}.

In this paper, we introduce a different approach to benchmarking gate--based devices with a simple
operational interpretation. In our method, we test how well the given device is at guessing which of
the two known von Neumann measurements were performed during the experiment. We implemented our
approach in an open-source Python library called PyQBench. The library supports any device available
through the Qiskit library, and thus can be used with providers such as IBM Q or Amazon Braket.
Along with the library, the PyQBench package contains a command line tool for running most common
benchmarking scenarios.

\section{Existing benchmarking methodologies and software}
Unsurprisingly, PyQBench is not the only software package for benchmarking gate--based devices.
While we believe that our approach has significant benefits over other benchmarking techniques, for
completeness, in this section we discuss some of the currently available similar software.

Probably the simplest benchmarking method one could devise is simply running known algorithms and
comparing outputs with the expected ones. Analyzing the frequency of the correct outputs, or the
deviation between actual and expected outputs distribution provides then a metric of the performance
of a given device.
 Libraries such as Munich Quantum Toolkit (MQT) \cite{mqt2022, mqt-bench}
or
 SupermarQ \cite{supermarq, supermarkq-github}
  contain
benchmarks leveraging multiple algorithms, such as Shor's algorithm or Grover's algorithm.
Despite being intuitive and easily interpretable, such benchmarks may have some problems. Most
importantly, they assess the usefulness of a quantum device only for a very particular algorithm,
and it might be hard to extrapolate their results to other algorithms and applications. For
instance, the inability of a device to consistently find factorizations using Shor's algorithms
does not tell anything about its usefulness in Variational Quantum Algorithm's.

Another possible approach to benchmarking quantum computers is randomized benchmarking. In this approach, one samples circuits to be run from some predefined set
of gates (e.g. from the Clifford group) and tests how much the output distribution obtained from the
device running these circuits differs from the ideal one. It is also common to concatenate randomly
chosen circuits with their inverses (which should yield the identity circuit) and run those
concatenated circuits on the device. Libraries implementing this approach include Qiskit \cite{qiskit-randomized} or PyQuil \cite{forest-benchmarking}.

Another quantity used for benchmarking NISQ devices is quantum volume. The quantum volume
characterizes capacity of a device for solving computational problems. It takes into account
multiple factors like number of qubits, connectivity and measurement errors. The Qiskit library allows one to
measure quantum volume of a device by using its \texttt{qiskit.ignis.verifica \\ tion.quantum\_volume}.
Other implementations of Quantum Volume can be found as well, see e.g. \cite{volume-in-practice}.


\section{Preliminaries and discrimination scheme approach}

In this section we describe how the benchmarking process in PyQBench works. To do so, we first
discuss necessary mathematical preliminaries. Then, we present the general form of the
discrimination scheme used in PyQBench and practical considerations on how to implement it taking
into account limitations of the current NISQ devices.

\subsection{Mathematical preliminaries}\label{sec:maths}

Let us first recall the definition of a von Neumann measurement, which is the only type of
measurement used in PyQBench. A von Neumann measurement $\PP$ is a collection of rank--one projectors
$\{\proj{u_0}, \ldots, \proj{u_{d-1}}\}$, called effects, that sum up to identity,
i.e. $ \, \, \sum_{i=0}^{d-1} \proj{u_i} = \1$. If $U$ is a unitary matrix of size $d$,
one can construct a von Neumann measurement $\PP_{U}$ by taking projectors onto its columns. In this
case we say that $\PP_{U}$ is described by the matrix $U$.

\begin{figure}[h!]
	\centering
	\includegraphics[scale=1.7]{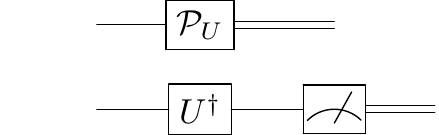}
	\caption{Implementation of a von Neumann measurement using measurement in computational basis.
	The upper circuit shows a symbolic representation of a von Neumann measurement $\PP_{U}$. The
	bottom, equivalent circuit depicts its decomposition into a change of basis followed
	by measurement in the $Z$ basis.}
	\label{fig:vonneumann}
\end{figure}

Typically, NISQ devices can only perform measurements in computational $Z$-basis, i.e. $U=\1$.
To implement an arbitrary von Neumann measurement $\PP_{U}$, one has to first apply $U^\dagger$
to the measured system and then follow with $Z$-basis measurement. This process, depicted in Fig.
\ref{fig:vonneumann}, can be viewed as performing a change of basis in which measurement is
performed prior to measurement in the computational basis.

\subsection{Discrimination scheme}\label{sec:discrimination-scheme}

Benchmarks in PyQBench work by experimentally determining the probability of correct discrimination
between two von Neumann measurements by the device under test and comparing the result with the
ideal, theoretical predictions.

Without loss of generality\footnote{Explaining why we can consider only discrimination scheme between $\PP_{\Id}$ and $\PP_{U}$ is beyond the scope of this paper. See \cite{puchala2018strategies} for a in depth explanation.}, we consider discrimination task between single qubit measurements
$\PP_\Id$, performed in the computational Z-basis, and an alternative measurement $\PP_U$ performed
in the basis $U$. Note, however, that the discrimination scheme described below can work
regardless of dimensionality of the system, see \cite{puchala2018strategies} for details.

\begin{figure}[h!]
	\centering
	\includegraphics[scale=1.7]{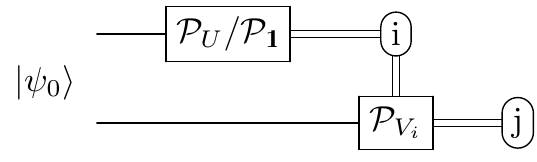}
	\caption{Theoretical  scheme of discrimination  between von Neumann measurements $\PP_{U}$ and $\PP_\Id$. }
	\label{fig:theoretical_scheme}
\end{figure}

In general, the discrimination scheme presented in Fig.~\ref{fig:theoretical_scheme}, requires an
auxiliary qubit. First, the joint system is prepared in some state $\ket{\psi_0}$. Then, one of the 
measurements,  either $\PP_U$ or $\PP_\1$, is performed on the first part of the system. Based on its
outcome $i$, we choose another POVM $\mathcal{P}_{V_i}$ and perform it on the second
qubit, obtaining the output in $j$. Finally, if $j=0$, we say that the performed measurement is
$\mathcal{P}_U$, otherwise we say that it was $\mathcal{P}_\Id$. Naturally, we need to repeat the
same procedure multiple times for both measurements to obtain a reliable estimate of the underlying
probability distribution. In PyQBench, we assume that the experiment is repeated the same number of
times for both $\PP_U$ and $\PP_\Id$.

Unsurprisingly, both the
$\ket{\psi_0}$ and the final measurements $\mathcal{P}_{V_i}$ have to be chosen specifically for
given $U$ to maximize the probability of a correct guess. The detailed description  how these
choices are made in \cite{watrous}, and for now we will focus
only how this scheme can be implemented on the actual devices, assuming that all the components are known.

\subsubsection{Implementation of discrimination scheme on actual NISQ devices}

Current NISQ devices are unable to perform conditional measurements, which is the biggest
obstacle to implementing our scheme on real hardware. However, we circumvent this problem by
slightly adjusting our scheme so that it only uses components available on current devices.
For this purpose, we use two possible options: using a postselection or a direct sum $V_0^\dagger\oplus V_1^\dagger$.

\begin{scheme}(Postselection)

	The first idea uses a postselection scheme. In the original scheme, we measure the first qubit
	and only then determine which measurement should be performed on the second one. Instead of
	doing this choice, we can run two circuits, one with $\PP_{V_0}$ and one with $\PP_{V_1}$ and
	measure both qubits. We then discard the results of the circuit for which label $i$ does not match
	measurement label $k$. Hence, the circuit for postselection looks as depicted in Fig.
	\ref{fig:postselection}.

	\begin{figure}[h!]
		\centering
		\includegraphics[scale=1.7]{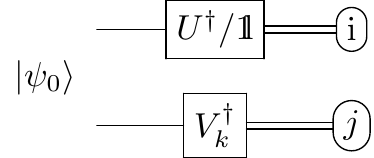}

		\caption{
			A schematic representation of the setup for distinguishing
			measurements $\PP_{U}$ and $\PP_{\Id}$ using postselection approach.
			In postselection scheme, one runs such circuits for both $k=0,1$ and discards results for cases when there is a mismatch between $k$ and $i$.
		}\label{fig:postselection}
	\end{figure}

	To perform the benchmark, one needs to run multiple copies of the postselection circuit,
	with both $\PP_U$ and $\PP_\1$. Each circuit has to be run in both variants, one with final
	measurement $\PP_{V_0}$ and the second with the final measurement $\PP_{V_1}$.
	The experiments can thus be grouped into classes identified by tuples of the form
	$(\mathcal{Q}, k, i, j)$, where $\mathcal{Q} \in \{\PP_U, \PP_\Id\}$ denotes the chosen
	measurement, $k \in \{0,1\}$ designates the final measurement used, and $i \in \{0,1\}$ and $j
	\in \{0,1\}$ being the labels of outcomes as presented in Fig. \ref{fig:postselection}.
	We then discard all the experiments for which $i \ne k$. The total number of valid experiments
	is thus:

	\begin{equation}
	N_\text{total} = \#\{(\QQ, k, i, j): k = i \}.
	\end{equation}

	Finally, we count the valid experiments resulting in successful discrimination.
	If we have chosen $\PP_U$, then we guess correctly iff $j=0$. Similarly, for
	$P_\Id$, we guess correctly iff $j=1$. If we define
	\begin{eqnarray}
	N_{\PP_U} &= \#\{(\mathcal{Q}, k, i, j): \mathcal{Q} = \PP_U, k = i, j = 0\}, \\
	N_{\PP_\Id} &= \#\{(\mathcal{Q}, k, i, j): \mathcal{Q} = \PP_\Id, k = i, j = 1\},
	\end{eqnarray}
	then the empirical success probability can be computed as

	\begin{equation}
	p_{\text{succ}}(\PP_{U}, \PP_{\Id}) = \frac{N_{\PP_U} + N_{\PP_\Id}}{N_{\text{total}}}.
	\end{equation}
	The $p_{\text{succ}}$ is the quantity reported to the user as the result of the benchmark.
\end{scheme}

\begin{scheme}(Direct sum)

The second idea uses the direct sum $V_0^\dagger \oplus V_1^\dagger$ implementation.     Here, instead of performing a conditional
measurement $\PP_{V_k}$, where $k\in \{0,1\}$,  we run circuits presented in
Fig.~\ref{fig:controlled}.

	\begin{figure}[h!]
		\centering
		\includegraphics[width=0.8\textwidth]{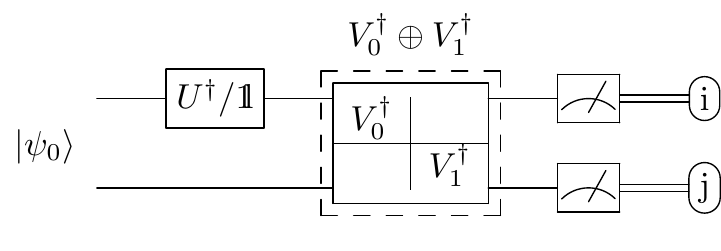}

		\caption{ A schematic representation of the setup for distinguishing
			measurements $\PP_{U}$ and $\PP_{\Id}$ using the $V_0^\dagger \oplus V_1^\dagger$ direct sum.
		}\label{fig:controlled}
	\end{figure}

	One can see why such a circuit is equivalent to the original discrimination scheme.
	If we rewrite the block-diagonal matrix $V_0^\dagger \oplus V_1^\dagger$ as follows:
	\begin{equation}
		\label{eq:directsum}
		V_0^\dagger \oplus V_1^\dagger = \proj{0}\otimes V_0^\dagger + \proj{1} \otimes V_1^\dagger,
	\end{equation}
	we can see that the direct sum in Eq. \eqref{eq:directsum} commutes with the measurement on the
	first qubit. Thanks to this, we can switch the order of operations to obtain the circuit from
	Fig. \ref{fig:directsum}. Now, depending on the outcome $i$, one of the summands in
	Eq. \eqref{eq:directsum} vanishes, and we end up performing exactly the same operations as in the
	original scheme.

	\begin{figure}[h!]
		\centering
		\includegraphics[width=\textwidth]{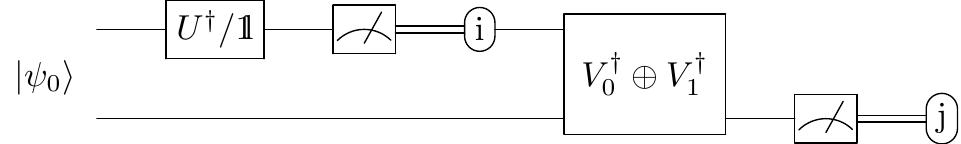}
		\caption{Rewritten representation of the setup for distinguishing
		measurements $\PP_{U}$ and $\PP_{\Id}$ using the $V_0^\dagger \oplus V_1^\dagger$ direct
		sum.
		}\label{fig:directsum}
	\end{figure}

	In this scheme, the experiment can be characterized by a pair $(\mathcal{Q}, i,j)$, where
	$\mathcal{Q} = \{ \PP_{U}, \PP_{\Id} \}$ and $i,j \in \{0,1\}$ are the output labels. The number of successful trials for $U$ and $\Id$,
	respectively, can be written  as
	\begin{eqnarray}
	N_{\PP_U} &= \#\{(\mathcal{Q},  i, j): \mathcal{Q} = \PP_U, j = 0\}, \\
	N_{\PP_\Id} &= \#\{(\mathcal{Q},  i, j): \mathcal{Q} = \PP_\Id, j = 1\}.
	\end{eqnarray}
	Then, the probability of correct discrimination between $\PP_{U} $ and $\PP_\Id$ is given by
	\begin{equation}
	p_{\text{succ}} = \frac{N_{\PP_{U}} + N_{\PP_{\Id}}}{N_{\text{total}}},
	\end{equation}
	where $N_{\text{total}}$ is the number of trials.
\end{scheme}

\subsubsection{Importance of choosing the optimal discrimination scheme}
In principle, the schemes described in the previous section could be used with any choice of
$\ket{\psi_0}$ and final measurements $\PP_{V_i}$. However, we argue that it is best to choose those
components in such a way that they maximize the probability of correct discrimination. To see that,
suppose that some choice of $\ket{\psi_0}, \PP_{V_0}, \PP_{V_1}$ yields the theoretical upper bound
of discriminating between two measurements of one, i.e. on a perfect quantum computer you will always
make a correct guess. Then, on real hardware, we might obtain any empirical value in range $\left[\frac{1}{2},
1\right]$. On the other hand, if we choose the components of our scheme such that the successful
discrimination probability is only $\frac{3}{5}$, the possible range of empirically obtainable probabilities
is only $\left[\frac{1}{2}, \frac{3}{5}\right]$. Hence, in the second case, the discrepancy between theoretical and empirical
results will be less pronounced.

\subsubsection{Constructing optimal discrimination scheme}
To construct the optimal discrimination scheme, one starts by calculating the probability of correct discrimination. Using the celebrated result by Helstrom~\cite{helstrom1976quantum}, one finds that the  optimal probability of correct discrimination between two quantum measurements, $\PP$  and $\mathcal{Q}$, is
\begin{equation}
p_{\text{succ}}(\PP, \mathcal{Q}) =  \frac12 + \frac14 \| \PP - \mathcal{Q} \|_\diamond,
\end{equation}
where
\begin{equation}
\|\PP - \QQ\|_\diamond = \max_{\| \ket{\psi}\|_1=1} \| \left( (\PP - \QQ) \otimes \1\right) (\proj{\psi}) \|_1.
\end{equation}
The quantum state $\ket{\psi_0}$ maximizing the diamond norm above is called the
\emph{discriminator}, and can be computed e.g. using semidefinite programming (SDP)
\cite{watrous,watrous2021simplier}. Furthermore,  using the proof of the Holevo-Helstrom theorem, it is
possible to construct corresponding unitaries $V_0$, $V_1$ to create the optimal
discrimination strategy. For brevity, we do not describe this procedure here. Instead, we refer the
interested reader to \cite{watrous}.

\section{Discrimination scheme for parameterized Fourier family and implementation}
\label{sec:fourier}
So far, we only discussed how the discrimination is performed assuming that all needed components
$\ket{\psi_0}$, $V_0$, and $V_1$ are known. In this section, we provide a concrete example using
parametrized Fourier family of measurements.

The parametrized Fourier family of measurements is defined as a set of the measurements
$\{\PP_{U_\phi}\colon \phi \in [0, 2\pi]\}$, where 
\begin{equation}
U_\phi = H
\left(\begin{array}{cc}1&0\\0&e^{i \phi}\end{array}\right)  H^\dagger,
\end{equation}
and $H$ is the Hadamard matrix of dimension two. For each element of this set, the discriminator is a Bell state:
\begin{equation}
\ket{\psi_{0}} = \frac{1}{\sqrt{2}} \left( \ket{00} + \ket{11} \right).
\end{equation}
Observe that $\ket{\psi_0}$ does not depend on the angle $\phi$. However, the unitaries $V_0$,
$V_1$ depend on $\phi$ and take the following form:
\begin{equation}
V_0 = \left(\begin{array}{cc}i \sin\left( \frac{\pi - \phi}{4} \right)&-i
\cos\left( \frac{\pi - \phi}{4} \right)\\ \cos\left( \frac{\pi -
	\phi}{4}\right)& \sin\left( \frac{\pi - \phi}{4} \right)\end{array}\right),
\end{equation}
\begin{equation}
V_1 = \left(\begin{array}{cc}-i \cos\left(\frac{\pi - \phi}{4}\right) &i
\sin\left( \frac{\pi - \phi}{4}\right)\\\sin\left( \frac{\pi - \phi}{4} \right)
&  \cos\left( \frac{\pi - \phi}{4} \right) \end{array}\right).
\end{equation}
Finally, the theoretical probability of correct discrimination between von Neumann
measurements $\PP_{U_\phi}$ and $\PP_{\Id}$ is given by
\begin{equation}
p_{\text{succ}}(\PP_{U_\phi}, \PP_{\Id}) = \frac{1}{2} + \frac{|1 - e^{i \phi}  |}{4} .
\end{equation}
We explore the construction of $\ket{\psi_0}$, $V_0$ and $V_1$ for parametrized Fourier family of measurements in
\ref{app:optimal-probability}.

 \section{Software description}
 \label{}
 This section is divided into two parts.
 In Section~\ref{sec:sortware-functionalities} we describe functionalities of PyQBench
 package. Next, in Section~\ref{sec:sortware-architecture}, we give a general overview of the
 software architecture.

\subsection{Software Functionalities}\label{sec:sortware-functionalities}

The PyQBench can be used in two modes: as a Python library and as a CLI script. When used as a
library, PyQBench allows the customization of discrimination scheme. The user provides
a unitary matrix $U$ defining the measurement to be discriminated, the discriminator $\ket{\psi_0}$,
and unitaries $V_0$ and $V_1$ describing the final measurement. The PyQBench library provides then
the following functionalities.

\begin{enumerate}
	\item Assembling circuits for both postselection and direct sum--based discrimination schemes.
	\item Executing the whole benchmarking scenario on specified backend (either real hardware or
	software simulator).
	\item Interpreting the obtained outputs in terms of discrimination probabilities.
\end{enumerate}
Note that the execution of circuits by PyQBench is optional. Instead, the user might want to opt in for
fine-grained control over the execution of the circuits. For instance, suppose the user wants to
simulate the discrimination experiment on a noisy simulator. In such a case, they can define
the necessary components and assemble the circuits using PyQBench. The circuits can then be altered,
e.g. to add noise to particular gates, and then run using any Qiskit backend by the user. Finally,
PyQBench can be used to interpret the measurements to obtain discrimination probability.

The PyQBench library also contains a readily available implementation of all necessary components
needed to run discrimination experiments for parametrized Fourier family of measurements, defined
previously in Section \ref{sec:fourier}. However, if one only wishes to use this particular family
of measurements in their benchmarks, then using PyQBench as a command line tool might be more straightforward. PyQBench's command line interface allows running the benchmarking process without
writing Python code. The configuration of CLI is done by YAML \cite{yaml} files describing the benchmark
to be performed and the description of the backend on which the benchmark should be run. Notably,
the YAML configuration files are reusable. The same benchmark can be used with different backends
and vice versa.

The following section describes important architectural decisions taken when creating PyQBench, and
how they affect the end-user experience.

\subsection{Software Architecture}\label{sec:sortware-architecture}

\subsubsection{Overview of the software structure}
As already described, PyQBench can be used both as a library and a CLI. Both functionalities are
implemented as a part of \texttt{qbench} Python package. The exposed CLI tool is also named
\texttt{qbench}. For brevity, we do not discuss the exact structure of the package here, and instead
refer an interested reader to the source code available at GitHub \cite{pyqbenchgithub} or at the reference manual
\cite{pyqbenchdocs}.

PyQBench can be installed from official Python Package Index (PyPI) by running \texttt{pip install
pyqbench}. In a properly configured Python environment the installation process should also make the
\texttt{qbench} command available to the user without a need for further configuration.

\subsubsection{Integration with hardware providers and software simulators}

PyQBench is built around the Qiskit \cite{qiskit} ecosystem. Hence, both the CLI tool and the
\texttt{qbench} library can use any Qiskit--compatible backend. This includes, IBM Q
backends (available by default in Qiskit) and Amazon Braket devices and simulators (available
through \texttt{qiskit-braket-provider} package \cite{qiskit-braket-provider, qiskit-braket-provider-github}).

When using PyQBench as library, instances of Qiskit backends can be passed to functions that expect
them as parameters. However, in CLI mode, the user has to provide a YAML file describing the
backend. An example of such file can be found in Section \ref{sec:examples}, and the detailed description of
the expected format can be found at PyQBench's documentation.

\subsubsection{Command Line Interface}
\label{sec:cli}

The Command Line Interface (CLI) of PyQBench has nested structure. The general form of the CLI
invocation is shown in listing \ref{lst:cli}.
\begin{lstlisting}[caption=Invocation of \texttt{qbench} script, label=lst:cli]
qbench <benchmark-type> <command> <parameters>
\end{lstlisting}
Currently, PyQBench's CLI supports only one type of benchmark (discrimination of parametrized
Fourier family of measurements), but we decided on structuring the CLI in a hierarchical
fashion to allow for future extensions. Thus, the only accepted value of \texttt{<benchmark-type>}
is \texttt{disc-fourier}.
The \texttt{qbench disc-fourier} command has four subcommands:

\begin{itemize}
	\item \texttt{benchmark}: run benchmarks. This creates either a result YAML file containing
	the measurements or an intermediate YAML file for asynchronous experiments.
	\item \texttt{status}: query status of experiments submitted for given benchmark. This command
	is only valid for asynchronous experiments.
	\item \texttt{resolve}: query the results of asynchronously submitted experiments and write the
	result YAML file. The output of this command is almost identical to the result obtained from
	synchronous experiments.
	\item \texttt{tabulate}: interpret the results of a benchmark and summarize them in the CSV file.
\end{itemize}
We present usage of each of the above commands later in section \ref{sec:examples}.

\subsubsection{Asynchronous vs. synchronous execution}
PyQBench's CLI can be used in synchronous and asynchronous modes. The
mode of execution is defined in the YAML file describing the backend (see Section
\ref{sec:examples} for an example of this configuration). We decided to couple the mode of execution to
the backend description because some backends cannot work in asynchronous mode.

When running \texttt{qbench disc-fourier benchmark} in asynchronous mode, the PyQBench submits all
the circuits needed to perform a benchmark and then writes an intermediate YAML file
containing metadata of submitted experiments. In particular, this metadata contains information on
correlating submitted job identifiers with particular circuits. The intermediate file can be
used to query the status of the submitted jobs or to resolve them, i.e. to wait for their completion
and get the measurement outcomes.

In synchronous mode, PyQBench first submits all jobs required to run the benchmark and then
immediately waits for their completion. The advantage of this approach is that no separate
invocation of \texttt{qbench} command is needed to actually download the measurement outcomes. The
downside, however, is that if the script is interrupted while the command is running, the
intermediate results will be lost. Therefore, we recommend using asynchronous mode
whenever possible.

\section{Illustrative examples}
\label{sec:examples}
In this section, we present two examples demonstrating the usage of PyQBench. In the first example,
we show how to implement a discrimination scheme for a user--defined measurement and possible ways
of using this scheme with \texttt{qbench} library. The second example demonstrates the usage of the
CLI. We show how to prepare the input files for the benchmark and how to run it using the
\texttt{qbench} tool.

\subsection{Using user-defined measurement with \texttt{qbench} package}
\begin{figure}[h!]
	\centering
	\includegraphics[scale=1.7]{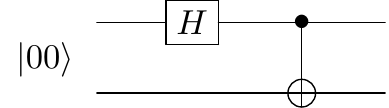}
	\caption{Decomposition of the Bell state $\ket{\psi_0}$. }
	\label{fig:discriminator}
\end{figure}

In this example, we will demonstrate how \texttt{qbench} package can be used with user--defined
measurement. For this purpose, we will use $U = H$ (the Hadamard gate). The detailed calculations
that lead to the particular form of the discriminator and final measurements can be found in
\ref{app:hadamard}.
\noindent
The explicit formula for discriminator in this example reads:
\begin{equation}
\ket{\psi_0} = \frac{1}{\sqrt{2}} (\ket{00} + \ket{11}),
\end{equation}
with final measurements being equal to
\begin{equation}
V_0 =
\left(\begin{array}{cc} \alpha & -\beta\\ \beta & \alpha \end{array}\right),
\end{equation}
and \begin{equation}
V_1 =
\left(\begin{array}{cc} -\beta & \alpha \\ \alpha & \beta \end{array}\right),
\end{equation}
where \begin{equation}
\alpha = \frac{\sqrt{2 - \sqrt{2}}}{2} = \cos\left( \frac{3}{8} \pi \right),
\end{equation}
\begin{equation}
\beta  = \frac{\sqrt{2  + \sqrt{2}}}{2} = \sin\left( \frac{3}{8} \pi \right).
\end{equation}
To use the above benchmarking scheme in PyQBench, we first need to construct circuits that can be executed by actual hardware. To this end, we need to represent each of the unitaries as a sequence of standard gates, keeping in mind that quantum circuits start execution from the  $\ket{00}$ state. The circuit taking $\ket{00}$ to the Bell state $\ket{\psi_0}$  comprises the Hadamard gate followed by CNOT gate on both qubits (see Fig.~\ref{fig:discriminator}).
For $V_0$ and $V_1$ observe that $V_0 = \operatorname{RY}\left(\frac{3}{4} \pi \right) $, where $\operatorname{RY}$ is rotation gate around the $Y$ axis defined by
\begin{equation}
\operatorname{RY}(\theta) =
\left(\begin{array}{cc} \cos\frac{\theta}{2} & -\sin\frac{\theta}{2} \\ \sin\frac{\theta}{2} & \cos\frac{\theta}{2} \end{array}\right).
\end{equation}
To obtain $V_1$  we need only to swap the columns, i.e.
\begin{equation}
V_1 = \operatorname{RY}\left(\frac{3}{4} \pi \right) \operatorname{X}.
\end{equation}
Finally, the optimal probability of correct discrimination is equal to
\begin{equation}
p_{\text{succ}}(\PP_{U}, \PP_{\Id}) = \frac{1}{2} + \frac{\sqrt{2}}{4}.
\end{equation}
We will now demonstrate how to implement this theoretical scheme in PyQBench. For this example we
will use the Qiskit Aer simulator \cite{aer}. First, we import the necessary functions and classes from
PyQBench and Qiskit. We also import \texttt{numpy} for the definition of \texttt{np.pi} constant and
the exponential function. The exact purpose of the imported functions will be described at the point
of their usage.
\begin{lstlisting}[language=Python, caption=Imports needed for running benchmarking example]
import numpy as np
from qiskit import QuantumCircuit, Aer
from qbench.schemes.postselection import benchmark_using_postselection
from qbench.schemes.direct_sum import benchmark_using_direct_sum
\end{lstlisting}
To implement the discrimination scheme in PyQBench, we need to define all the necessary components
as Qiskit instructions. We can do so by constructing a circuit object acting on qubits 0 and 1 and
then converting them using \texttt{to\_instruction()} method.
\begin{lstlisting}[language=Python, caption= Defining components for Hadamard experiment]
def state_prep():
	circuit = QuantumCircuit(2)
	circuit.h(0)
	circuit.cnot(0, 1)
	return circuit.to_instruction()

def u_dag():
	circuit = QuantumCircuit(1)
	circuit.h(0)
	return circuit.to_instruction()

def v0_dag():
	circuit = QuantumCircuit(1)
	circuit.ry(-np.pi * 3 / 4, 0)
	return circuit.to_instruction()

def v1_dag():
	circuit = QuantumCircuit(1)
	circuit.ry(-np.pi * 3 / 4, 0)
	circuit.x(0)
	return circuit.to_instruction()

def v0_v1_direct_sum_dag():
	circuit = QuantumCircuit(2)
	circuit.ry(-np.pi * 3 / 4, 0)
	circuit.cnot(0, 1)
	return circuit.to_instruction()
\end{lstlisting}
We now construct a backend object, which in this case is an instance of Aer simulator.

\begin{lstlisting}[language=Python, caption=Defining a backend]
simulator = Aer.get_backend("aer_simulator")
\end{lstlisting}

In the simplest scenario, when one does not want to tweak execution details and simply wishes to run
the experiment on a given backend, everything that is required is now to run
\texttt{benchmark\_using\_postselection} or \texttt{benchmark\_using\_direct\_sum} function,
depending on the user preference.

\begin{lstlisting}[language=Python, caption=Simulation benchmark by using postselection]
postselection_result = benchmark_using_postselection(
	backend=simulator,
	target=0,
	ancilla=1,
	state_preparation=state_prep(),
	u_dag=u_dag(),
	v0_dag=v0_dag(),
	v1_dag=v1_dag(),
	num_shots_per_measurement=10000,
)
\end{lstlisting}

\begin{lstlisting}[language=Python, caption=Simulation benchmark by using direct sum]
direct_sum_result = benchmark_using_direct_sum(
	backend=simulator,
	target=1,
	ancilla=2,
	state_preparation=state_prep(),
	u_dag=u_dag(),
	v0_v1_direct_sum_dag=v0_v1_direct_sum_dag(),
	num_shots_per_measurement=10000,
)
\end{lstlisting}
The \texttt{postselection\_result} and  \texttt{direct\_sum\_result} variables contain now the
empirical probabilities of correct discrimination. We can compare them to the theoretical value and
compute the absolute error.

\begin{lstlisting}[language=Python, caption=Examining the benchmark results]
p_succ = (2 + np.sqrt(2)) / 4
print(f"Analytical p_succ = {p_succ}")
print(
f"Postselection: p_succ = {postselection_result}, abs. error = {p_succ - postselection_result}"
)
print(f"Direct sum: p_succ = {direct_sum_result}, abs. error = {p_succ - direct_sum_result}")
\end{lstlisting}

\begin{lstlisting}
Analytical p_succ = 0.8535533905932737
Postselection: p_succ = 0.8559797193791593, abs. error = -0.0024263287858855564
Direct sum: p_succ = 0.85605, abs. error = -0.0024966094067262468
\end{lstlisting}

\begin{figure}[ht!]
	\centering
	\begin{subfigure}[b]{0.9\textwidth}
		\centering
		\includegraphics[width=0.8\textwidth]{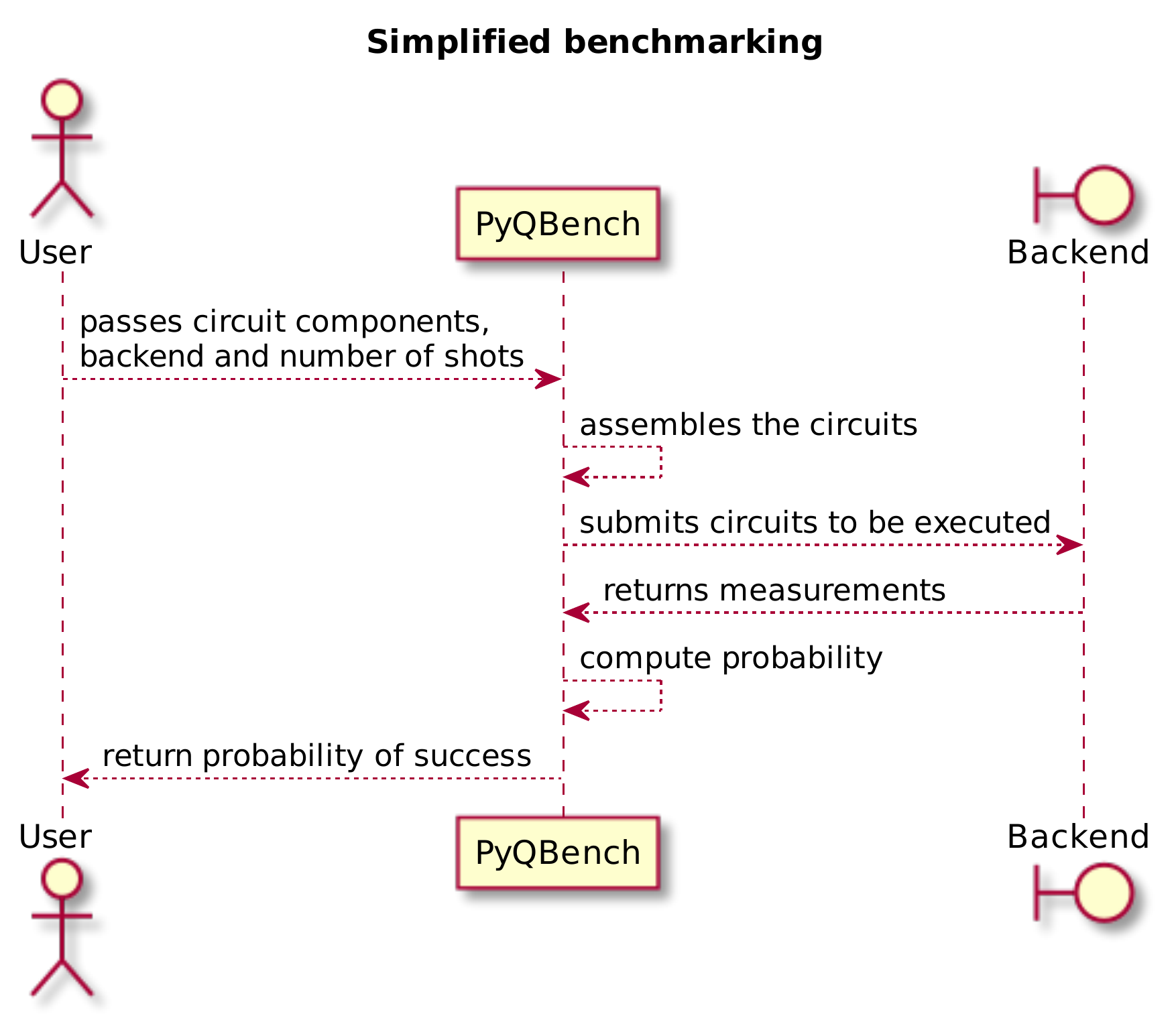}
		\label{fig:simplified}
	\end{subfigure}
	\vfill
	\begin{subfigure}[b]{0.9\textwidth}
		\centering
	\includegraphics[width=0.8\textwidth]{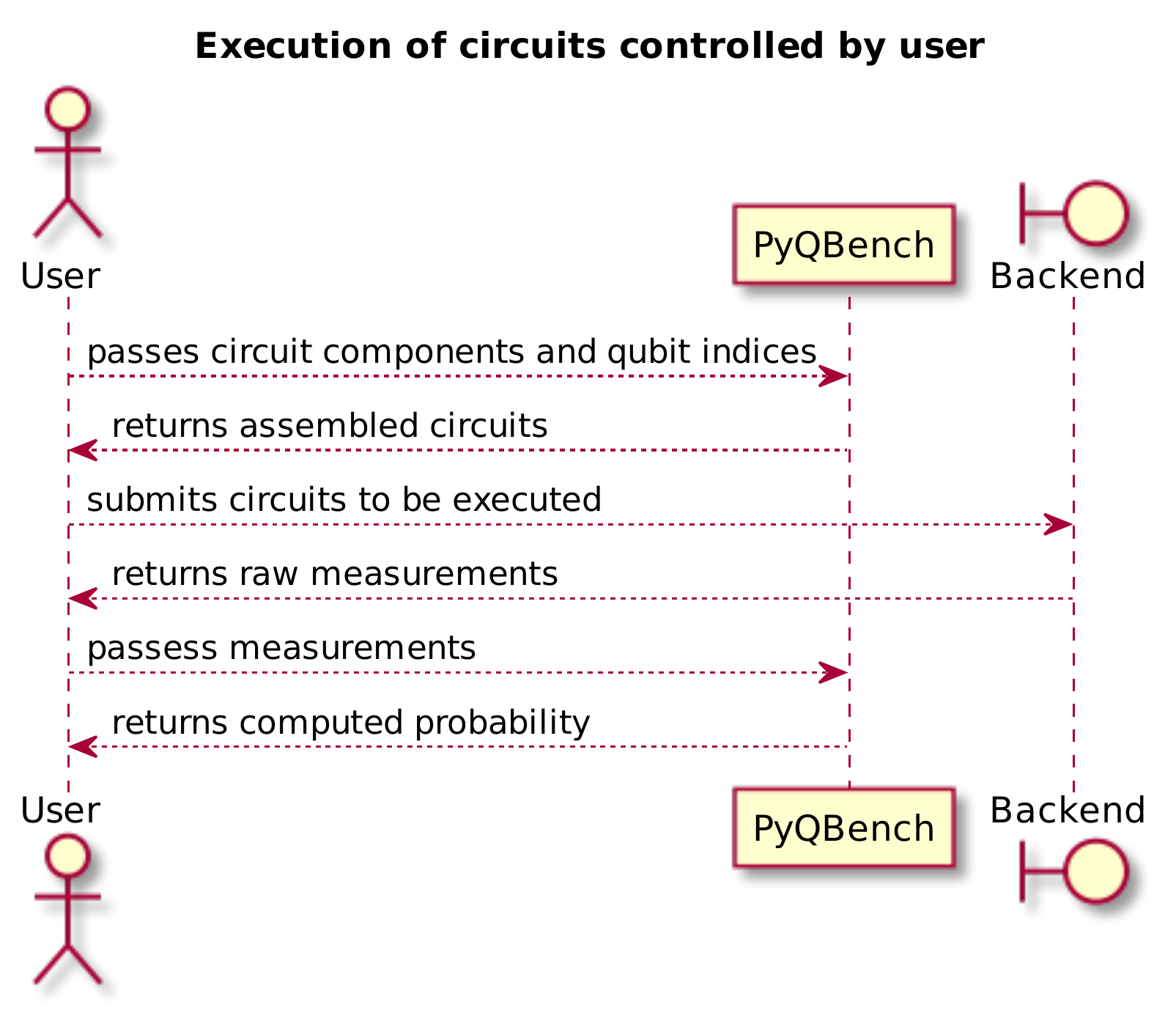}
		\label{fig:execution}
	\end{subfigure}
	\caption{Differences between simplified (top) and user--controlled (bottom) execution of
	benchmarks in PyQBench. Compared to simplified benchmarking, in user-controlled benchmarks the
	user has direct access to the circuits being run, and hence can alter them (e.g. by adding
	noise) and/or choose the parameters used for executing them on the backend.}
	\label{fig:diagrams}
\end{figure}

In the example presented above we used functions that automate the whole process -- from the circuit assembly, through running the simulations to interpreting the results. But what if we want more control over some parts of this process?
One possibility would be to add some additional parameters to \texttt{benchmark\_using\_xyz}
functions, but this approach is not scalable. Moreover, anticipating all possible uses cases
isimpossible. Therefore, we decided on another approach. PyQBench provides functions performing:
\begin{enumerate}
\item Assembly of circuits needed for experiment, provided the components discussed above.
\item Interpretation of the obtained measurements.
\end{enumerate}
 The difference between the two approaches is illustrated on the diagrams in Fig. \ref{fig:diagrams}.

For the rest of this example we focus only on the postselection case, as the direct sum case is
analogous. We continue by importing two more functions from PyQBench.
\begin{lstlisting}[language=Python, caption= Assembling circuits]
from qbench.schemes.postselection import (
    assemble_postselection_circuits,
    compute_probabilities_from_postselection_measurements,
)

circuits = assemble_postselection_circuits(
	target=0,
	ancilla=1,
	state_preparation=state_prep(),
	u_dag=u_dag(),
	v0_dag=v0_dag(),
	v1_dag=v1_dag(),
)

\end{lstlisting}
Recall that for a postselection scheme we have two possible choices of the "unknown" measurement and
two possible choices of a final measurement, which gives a total of four circuits needed to run the
benchmark. The function \texttt{assemble\_postselection\_circuits} creates all four circuits and
places them in a dictionary with keys \texttt{"id\_v0"}, \texttt{"id\_v1"}, \texttt{"u\_v0"},
\texttt{"u\_v1"}.

We will now run our circuits using noisy and noiseless simulation. We start by creating a noise
model using Qiskit.

\begin{lstlisting}[language=Python, caption=Adding noise model]
from qiskit.providers.aer import noise

error = noise.ReadoutError([[0.75, 0.25], [0.8, 0.2]])

noise_model = noise.NoiseModel()
noise_model.add_readout_error(error, [0])
noise_model.add_readout_error(error, [1])
\end{lstlisting}
Once we have our noise model ready, we can execute the circuits with and without noise. To this end,
we will use Qiskit’s \texttt{execute} function. One caveat is that we have to keep track which
measurements correspond to which circuit. We do so by fixing an ordering on the keys in the \texttt{circuits} dictionary.
\begin{lstlisting}[language=Python, caption=Running  circuits]
from qiskit import execute

keys_ordering = ["id_v0", "id_v1", "u_v0", "u_v1"]
all_circuits = [circuits[key] for key in keys_ordering]

counts_noisy = execute(
	all_circuits,
	backend=simulator,
	noise_model=noise_model,
	shots=10000
).result().get_counts()

counts_noiseless = execute(
	all_circuits,
	backend=simulator,
	shots=10000
).result().get_counts()
\end{lstlisting}
Finally, we use the measurement counts to compute discrimination probabilities using \texttt{compute\_probabilities\_from\_postselection\_measurements} function.

\begin{lstlisting}[language=Python, caption=Computation probabilities]
prob_succ_noiseless = compute_probabilities_from_postselection_measurements(
	id_v0_counts=counts_noiseless[0],
	id_v1_counts=counts_noiseless[1],
	u_v0_counts=counts_noiseless[2],
	u_v1_counts=counts_noiseless[3],
)


prob_succ_noisy = compute_probabilities_from_postselection_measurements(
	id_v0_counts=counts_noisy[0],
	id_v1_counts=counts_noisy[1],
	u_v0_counts=counts_noisy[2],
	u_v1_counts=counts_noisy[3],
)

\end{lstlisting}

We can now examine the results. As an example, in one of our runs, we obtained
\texttt{prob\_succ\_noiseless = 0.8524401115559386} and \texttt{prob\_succ\_noisy =
0.5017958400693446}. As expected, for noisy simulations, the result lies further
away from the target value of \texttt{0.8535533905932737}.

This concludes our example. In the next section, we will show how to use PyQBench's CLI.

\subsection{Using \texttt{qbench} CLI}
Using PyQBench as a library allows one to conduct a two--qubits benchmark with arbitrary von Neumann
measurement. However, as discussed in the previous guide, it requires writing some amount of code.
For a Fourier parametrized family of measurements, PyQBench offers a simplified way of conducting
benchmarks using a Command Line Interface (CLI). The workflow with PyQBench's CLI can be summarized
as the following list of steps::

\begin{enumerate}

\item Preparing configuration files describing the backend and the experiment scenario.
\item Submitting/running experiments. Depending on the experiment scenario, execution can be synchronous, or asynchronous.
\item (optional) Checking the status of the submitted jobs if the execution is asynchronous.
\item Resolving asynchronous jobs into the actual measurement outcomes.
\item Converting obtained measurement outcomes into tabulated form.
\end{enumerate}
\subsubsection{Preparing configuration files}
The configuration of PyQBench CLI is driven by YAML files. The first configuration file
describes the experiment scenario to be executed. The second file describes the backend.
Typically, this backend will correspond to the physical device to be benchmarked, but for testing
purposes one might as well use any other Qiskit--compatible backend including simulators.
Let us first describe the experiment configuration file, which might look as follow.
\begin{lstlisting}[language=Python, caption=Defining the experiment, label=lst:experiment]
type: discrimination-fourier
qubits:
	- target: 0
	  ancilla: 1
	- target: 1
      ancilla: 2
angles:
	start: 0
	stop: 2 * pi
	num_steps: 3
gateset: ibmq
method: direct_sum
num_shots: 100
\end{lstlisting}
The experiment file contains the following fields:
\begin{itemize}
	\item \texttt{type}: a string describing the type of the experiment. Currently, the only option of  \texttt{type} is \texttt{discrimination-fourier}.
	\item \texttt{qubits}: a list enumerating pairs of qubits on which the experiment should be run.
	For configuration in listing \ref{lst:experiment}, the benchmark will run on two pairs of qubits. The first pair is 0 and 1, and
	the second one is 1 and 2. We decided to describe a pair by using \texttt{target} and
	\texttt{ancilla} keys rather than using a plain list to emphasize that the role of qubits in the
	experiment is not symmetric.
	\item \texttt{angles}: an object describing the range of angles for Fourier parameterized
	family. The described range is always uniform, starts at the \texttt{start}, ends at
	\texttt{stop} and contains \texttt{num\_steps} points, including both \texttt{start} and
	\texttt{stop}. The \texttt{start} and \texttt{stop} can be arithmetic expressions using
	\texttt{pi} literal. For instance, the range defined in listing \ref{lst:experiment} contains
	three points: 0, $\pi$ and $2\pi$.
 	 \item \texttt{gateset}: a string describing the set of gates used in the decomposition of
 	 circuits in the experiment. The PyQBench contains explicit implementations of circuits The
 	 possible options are \texttt{[ibmq, lucy, rigetti]}, corresponding to decompositions
 	 compatible with IBM Q devices, OQC Lucy device, and Rigetti devices. Alternatively, one might
 	 wish to turn off the decomposition by using a special value \texttt{generic}. However, for this
 	 to work a backend used for the experiment must natively implement all the gates needed for the
 	 experiment, as described in \ref{sec:fourier}.
 	\item \texttt{method}: a string, either \texttt{postselection} or \texttt{direct\_sum}
 	determining which implementation of the conditional measurement is used.
 	\item \texttt{num\_shots}: an integer defines how many shots are performed in the experiment for
 	a particular angle, qubit pair and circuit. Note that if one wishes to compute the total number
 	of shots in the experiment, it is necessary to take into account that the \texttt{postselection}
 	method uses twice as many circuits as the \texttt{direct\_sum} method.
\end{itemize}

The second configuration file describes the backend. We decided to decouple the
experiment and the backend files because it facilitates their reuse. For instance, the same
experiment file can be used to run benchmarks on multiple backends, and the same backend description
file can be used with multiple experiments.

Different Qiskit backends typically require different data for their initialization. Hence, there
are multiple possible formats of the backend configuration files understood by PyQBench. We refer
the interested reader to the PyQBench's documentation. Below we describe an example YAML file
describing IBM Q backend named
Quito.
\begin{lstlisting}[language=Python, caption=IBMQ backend, label=lst:backend]
name: ibmq_quito
asynchronous: false
provider:
	hub: ibm-q
	group: open
	project: main
\end{lstlisting}
IBMQ backends typically require an access token to IBM Quantum Experience. Since it would be unsafe
to store it in plain text, the token has to be configured separately in \texttt{IBMQ\_TOKEN}
environmental variable.

\subsubsection{Remarks on using the asynchronous flag}
For backends supporting asynchronous execution, the \texttt{asynchronous} setting can be configured
to toggle it.
For asynchronous execution to work, the following conditions have to be met:
\begin{itemize}
\item Jobs returned by the backend have unique \texttt{job\_id}.
\item Jobs are retrievable from the backend using the \texttt{backend.retrieve\_job} method, even from another process (e.g. if the original process running the experiment has finished).
\end{itemize}
Since PyQBench cannot determine if the job retrieval works for a given backend, it is the user's
responsibility to ensure that this is the case before setting \texttt{asynchronous} to \texttt{true}.

\subsubsection{Running the experiment and collecting measurements data}
After preparing YAML files defining experiment and backend,
running the benchmark can be launched by using the following command line invocation:
\begin{lstlisting}[language=Python]
qbench disc-fourier benchmark experiment_file.yml backend_file.yml
\end{lstlisting}
The output file will be printed to stdout. Optionally, the \texttt{- -output OUTPUT} parameter might be provided to write the output to the \texttt{OUTPUT} file instead.
\begin{lstlisting}[language=Python]
qbench disc-fourier benchmark experiment_file.yml backend_file.yml --output async_results.yml
\end{lstlisting}
The result of running the above command can be twofold:
\begin{itemize}
	\item If backend is asynchronous, the output will contain intermediate data containing, amongst others, \texttt{job\_ids} correlated with the circuit they correspond to.
	\item If the backend is synchronous, the output will contain measurement outcomes (bitstrings) for each of the circuits run.
\end{itemize}

For synchronous experiment, the part of output looks similar to the one below. The whole YAML file can be seen in \ref{app:example}.
\begin{lstlisting}[language=Python]
data:
- target: 0
  ancilla: 1
  phi: 0.0
  results_per_circuit:
  - name: id
  histogram: {'00': 28, '01': 26, '10': 21, '11': 25}
  mitigation_info:
	target: {prob_meas0_prep1: 0.052200000000000024, prob_meas1_prep0: 0.0172}
	ancilla: {prob_meas0_prep1: 0.05900000000000005, prob_meas1_prep0: 0.0202}
  mitigated_histogram: {'00': 0.2637212373658018, '01': 0.25865061319892463, '10': 0.2067279352110304, '11': 0.2709002142242433}
\end{lstlisting}
	The data includes \texttt{target, ancilla, phi},
	and \texttt{results\_per\_circuit}.   The first three pieces of information have already been described. The last data  \texttt{results\_per\_circuit} gives us
	the following additional information:
	\begin{itemize}
		\item  \texttt{name}:  the information which measurement is used during experiment, either
		string \texttt{"u"} for $\PP_{U}$ or string \texttt{"id"} for $\PP_{\Id}$. In this example
		we consider $\PP_{\Id}$.
		\item \texttt{histogram}: the dictionary with measurements' outcomes. The keys represent
		possible bitstrings, whereas the values are the number of occurrences.
		\item \texttt{mitigation\_info}:  for some backends (notably for backends corresponding to
		IBM Q devices), \texttt{backends.properties().qubits} contains information that might be
		used for error mitigation using the MThree method \cite{mthree, mthreepublication}. If this
		info is available, it will be stored in the \texttt{mitigation\_info} field, otherwise this field will be absent.
		\item \texttt{mitigated\_histogram}: the histogram with measurements' outcomes after the error mitigation.
	\end{itemize}

\subsubsection{(Optional) Getting status of asynchronous jobs}
 PyQBench provides also a helper command that will fetch the statuses of asynchronous jobs. The command is:
\begin{lstlisting}[language=Python]
qbench disc-fourier status async_results.yml
\end{lstlisting}
and it will display dictionary with histogram of statuses.

\subsubsection{Resolving asynchronous jobs}
For asynchronous experiments, the stored intermediate data has to be resolved in actual
measurements' outcomes. The following command will wait until all jobs are completed and then write
a result file.
\begin{lstlisting}[language=Python]
qbench disc-fourier resolve async-results.yml resolved.yml
\end{lstlisting}
The resolved results, stored in \texttt{resolved.yml}, would look just like if the experiment was
run synchronously. Therefore, the final results will look the same no matter in which mode the
benchmark was run, and hence in both cases the final output file is suitable for being an input for
the command computing the discrimination probabilities.

\subsubsection{Computing probabilities}
As a last step in the processing workflow, the results file has to be passed to \texttt{tabulate}
command:
\begin{lstlisting}[language=Python]
qbench disc-fourier tabulate results.yml results.csv
\end{lstlisting}
A sample CSV file is provided below:
\begin{table}[ht!]
\begin{center}
	\begin{tabular}{|c c c c c c|} 
		\hline
		target & ancilla & phi & ideal$\_$prob  & disc$\_$prob & mit$\_$disc$\_$prob \\ [0.5ex] 
		\hline\hline
		0 & 1 & 0 & 0.5 & 0.46 & 0.45  \\ 
		\hline
		0 & 1 & 3.14 & 1 & 0.95 & 0.98  \\
		\hline
		0 & 1 & 6.28 & 0.5 & 0.57 & 0.58  \\
		\hline
		1 & 2 & 0 & 0.5  & 0.57 & 0.57 \\
		\hline
		1 & 2 & 3.14 & 1 & 0.88 & 0.94  \\ 
		\hline
		1 & 2 & 6.28 & 0.5 & 0.55 & 0.56  \\ 
		\hline
	\end{tabular}
\caption{The resulting CSV file contains table with columns \texttt{target}, \texttt{ancilla}, \texttt{phi},
	\texttt{ideal\_prob}, \texttt{disc\_prob} and, optionally, \texttt{mit\_disc\_prob}. Each row in the
	table describes results for a tuple of \texttt{(target, ancilla, phi)}.  The reference optimal value
	of discrimination probability is present in \texttt{ideal\_prob} column, whereas the obtained,
	empirical discrimination probability can be found in the \texttt{disc\_prob} column. The
	\texttt{mit\_disc\_prob} column contains empirical discrimination probability after applying the
	\texttt{Mthree} error mitigation \cite{mthree, mthreepublication}, if it was applied.}
\label{fig:tabulateresults}
\end{center}
\end{table}

\section{Impact}
With the surge of availability of quantum computing architectures in recent
years it becomes increasingly difficult to keep track of their relative
performance. To make this case even more difficult, various providers give
access to different figures of merit for their architectures. Our package allows
the user to test various architectures, available through \texttt{qiskit} and
Amazon BraKet using problems with simple operational interpretation. We provide
one example built-in in the package. Furthermore, we provide a powerful tool for
the users to extend the range of available problems in a way that suits their
needs. 

Due to this possibility of extension, the users are able to test specific
aspects of their architecture of interest. For example, if their problem is
related to the amount of coherence (the sum of absolute value of off-diagonal
elements) of the states present during computation, they are able to quickly
prepare a custom experiment, launch it on desired architectures, gather the
result, based on which they can decide which specific architecture they should
use.

Finally, we provide the source code of PyQBench on GitHub~\cite{pyqbenchgithub} under an open
source license which will allow users to utilize and extend our package in their
specific applications.

\section{Conclusions}
\label{}

In this study, we develop  a Python library PyQBench, an innovative open-source framework for benchmarking
gate-based quantum computers.

PyQBench can benchmark NISQ devices by verifying their capability of
discriminating between two von Neumann measurements. PyQBench offers a simplified, ready-to-use,
command line interface (CLI) for running benchmarks using a predefined parameterized Fourier
family of measurements. For more advanced scenarios, PyQBench offers a way of employing user-defined
measurements instead of predefined ones.
\section{Conflict of Interest}
We wish to confirm that there are no known
conflicts of interest associated with this publication and there has been no
significant financial support for this work that could have influenced its out-
come.

\section*{Acknowledgements}

This work is  supported by
the project “Near-term quantum computers Challenges, optimal implementations and applications” under Grant Number POIR.04.04.00-00-17C1/18-00, which is carried out within the Team-Net programme of the Foundation for Polish Science co-financed by the European Union under the European Regional Development Fund.
PL is also a holder of European Union scholarship through the European Social Fund,
grant InterPOWER (POWR.03.05.00-00-Z305).

\bibliographystyle{unsrt}
\bibliography{references}

\appendix

\section{Mathematical preliminaries} \label{app:preliminaries}

Let $\mathcal{M}_{d_1,d_2}$ be the set of all matrices of dimension $d_1 \times d_2$ over
the field $\mathbb{C}$. For  simplicity, square matrices will be denoted by
$\mathcal{M}_d$.
By $\Omega_d$, we will denote the set of quantum states, that is
positive semidefinite operators having trace equal to one.
The subset of $\mathcal{M}_d$ consisting of unitary matrices will be denoted
by $\UU_d$, while its subgroup of diagonal unitary operators will be denoted by
$\DD \UU_d$.

We will also need a linear mapping transforming $\mathcal{M}_{d_1}$ into
$\mathcal{M}_{d_2}$, which will be denoted
\begin{equation}
\Phi: \mathcal{M}_{d_1 } \rightarrow \mathcal{M}_{d_2}.
\end{equation}
There
exists a bijection between the set of linear mappings $\Phi$ and the set of matrices $\mathcal{M}_{d_1d_2}$,  known as the Choi-Jamio{\l}kowski isomorphism.
For a given linear mapping $\Phi$ the corresponding Choi operator $J(\Phi)$ is explicitly written as
\begin{equation}
J(\Phi) \coloneqq \sum_{i,j=0}^{d- 1} \Phi(\ketbra{i}{j}) \otimes \ketbra{i}{j}. \end{equation}

We also introduce a special subset of all mappings $\Phi$, called quantum channels, which are completely positive
and trace preserving (CPTP).
In this work we will consider a special class of quantum channels, called unitary channels.  A
quantum channel
$\Phi_{U}$ is said to be a unitary channel if it has the following form $\Phi_U(\cdot) = U \cdot U^\dagger$ for any $U \in
\UU_d$.

Let us recall a  general form of a quantum measurement, so called Positive Operator Valued
Measure (POVM). A POVM $\PP$ is a collection of positive semidefinite operators $\{E_1, \ldots, E_m
\}$, called effects, that sum up to the identity operator, \ie $ \, \, \sum_{i=1}^m E_i = \1$.
In PyQBench, we are interested only in von Neumann measurements, that is measurements
for which all the effects are rank-one projectors. Every such measurement can be
parameterized by a unitary matrix $U \in \mathcal{U}_d$ with effects $\{\proj{u_0}, \ldots, \proj{u_{d-1}}\}$,
are created by taking $\ket{u_i}$ as  $(i+1)$-th column of the unitary matrix $U$.
We will denote von Neumann measurements described by the matrix $U$ by $\PP_{U}$.
The action of
von Neumann measurement $\PP_{U}$ on some state $\rho \in \Omega_d$ can be
seen as  a measure-and-prepare quantum channel as follows \begin{equation}
\PP_{U} : \rho \rightarrow \sum_{i=0}^{d-1} \bra{u_i} \rho \ket{u_i} \proj{i}.
\end{equation}
Moreover, observe that each von Neumann measurement $\PP_{U}$ poses a  composition of a unitary channel $\Phi_{U^\dagger}$ and the maximally dephasing channel $\Delta$, that means $\PP_{U} = \Delta \circ \Phi_{U^\dagger}$.

We need to also briefly discuss about the distance between quantum operations. From \cite[Theorem 1]{puchala2018strategies}, the distance between measurements $\PP_U$ and
$\PP_\Id$ can be expressed in the notion of diamond norm, that is
\begin{equation}
\|\PP_U - \PP_\Id\|_\diamond = \min_{E \in \diaguni_d} \|\Phi_{UE} -
\Phi_\Id\|_\diamond.
\end{equation} To express the distance between unitary channels, we need to introduce the definition of numerical range \cite{numericalrangle}. The set \begin{equation}
W(A) =\{\bra{x}A\ket{x}: \ket{x} \in
\mathbb{C}^d, \;
\;\braket{x}{x}=1\}
\end{equation}
is called the numerical range of  a given matrix $A \in \mathcal{M}_d$.
The detailed properties of the numerical range and its generalizations we can read on the website~\cite{nr}.

Due to the definition of $W(A)$, the distance between two unitary channels $\Phi_{U} $ and $\Phi_\Id$
can be written as
\begin{equation}
\| \Phi_U  - \Phi_{\1} \|_\diamond = 2 \sqrt{1-\nu^2},
\end{equation}
where $\nu = \min_{x \in W(U^\dagger)} |x|  $.

\section{Discrimination task for Hadamard gate}\label{app:hadamard}
For the discrimination task between von Neumann measurements $\PP_{U}$ and $\PP_\Id$, where $U = H$ (the Hadamard gate),  the key is to calculate the diamond norm $\| \mathcal{P}_H - \PP_\Id \|_\diamond$ and
determine the discriminator $\ket{\psi_0}$.
Using semidefinite programming \cite{watrous2021simplier}, we obtain
\begin{equation}
\| \mathcal{P}_H - \PP_\Id \|_\diamond = \sqrt{2}.
\end{equation}
From \cite{lewandowska2021certification} we have
\begin{equation}
\| \PP_{H} - \PP_{\Id} \|_\diamond = \| \Phi_{HE_0} - \Phi_{\Id}\|_\diamond,
\end{equation}
where $\Phi_{U}$ is a unitary channel and
 $E_0$
is of the form
\begin{equation}
E_0 = \frac{1}{\sqrt{2}} \left( \begin{matrix}
1 + i & 0  \\  0 & -1-i
\end{matrix} \right).
\end{equation}
Next, in order to construct the discriminator $\ket{\psi_0}$ we use Lemma 5 and the proof of Theorem 1 in \cite{puchala2018strategies}. We show that there exist states
 $\rho_1 $  and $\rho_2$ of the form $ \rho_1  = \frac{1}{2} \left( \begin{matrix}
 1  & i  \\  -i & 1
 \end{matrix} \right)  $  and $\rho_2 = \frac{1}{2} \left( \begin{matrix}
 1 & -i \\  i & 1
 \end{matrix} \right), $ respectively. Thus, we construct the quantum state $\rho_0$ as follows:
 \begin{equation}
 \rho_0 = \frac{1}{2} \rho_1 + \frac{1}{2} \rho_2 = \frac{1}{2} \left( \begin{matrix}
 1  &  0  \\  0  & 1
 \end{matrix} \right).
 \end{equation}
 According to the Lemma 5 and the proof of Theorem 1 in \cite{puchala2018strategies} we assume that \begin{equation}
 \ket{\psi_0} = \left.\left| \sqrt{\rho_0^\top}\right\rangle \right\rangle.
 \end{equation}
 It directly implies that
 \begin{equation}
 \ket{\psi_0} = \frac{1}{\sqrt{2}}(\ket{00} + \ket{11}).  \end{equation}
 Next, from Holevo-Helstrom theorem~\cite{watrous}, we determine the final measurement $\PP_{V_i}$.
 Let us consider \begin{equation}
 X = \left(\PP_{H} \otimes \Id \right)(\proj{\psi_0}) - \left(\PP_{\Id} \otimes \Id \right) (\proj{\psi_0}).
  \end{equation}
 From the Hahn-Jordan decomposition \cite{watrous}, let us note
 \begin{equation}
 X = P - Q,
 \end{equation}
 where $P, Q \ge 0 $.
 Let us define projectors $\Pi_P$ and $\Pi_Q$ onto  $\text{im}(P)$ and $\text{im}(Q)$,
 respectively. Observe, that $P $ and $Q$ are block-diagonal.  Then,  $\Pi_P$ and $\Pi_Q$ have the following forms
 \begin{equation}
 \Pi_P = \left(\begin{array}{cc}\proj{x_p}&0\\0&\proj{y_p}\end{array}\right),
 \end{equation}
 and
 \begin{equation}
 \Pi_Q = \left(\begin{array}{cc}\proj{x_q}&0\\0&\proj{y_q}\end{array}\right).
 \end{equation}
 Hence, we define $V_0$ as
 \begin{equation}
 \begin{cases} \ket{x_p} =  V_0  \ket{0} \\  \ket{x_q} =  V_0 \ket{1} \end{cases}
 \end{equation}
 and $V_1$ as
 \begin{equation}
 \begin{cases}
 \ket{y_p} =   V_1 \ket{0} \\
  \ket{y_q} = V_1  \ket{1}
 \end{cases}.
 \end{equation}
 For the discrimination task between $\PP_{H}$ and $\PP_{\Id}$ the explicit form of $V_0$ and $V_1$ is given as follows (see also \texttt{mathematics/optimal\_final\_measurement\_ \\ discrimination.nb} in the source code repository):
 \begin{equation}
 V_0 =
 \left(\begin{array}{cc} \alpha & -\beta\\ \beta & \alpha \end{array}\right),
 \end{equation}
 and \begin{equation}
 V_1 =
 \left(\begin{array}{cc} -\beta & \alpha \\ \alpha & \beta \end{array}\right),
 \end{equation}
 where \begin{equation}
 \alpha = \frac{\sqrt{2 - \sqrt{2}}}{2} = \cos\left( \frac{3}{8} \pi \right),
 \end{equation}
 and
 \begin{equation}
 \beta  = \frac{\sqrt{2  + \sqrt{2}}}{2} = \sin\left( \frac{3}{8} \pi \right).
 \end{equation}

\section{Optimal probability for parameterized Fourier family} \label{app:optimal-probability}
Let us focus on single-qubit von Neumann measurements $\PP_\1$ and $\PP_U$.
Assume that the unitary matrix $U$ is of the form
\begin{equation}
U = H
\left(\begin{array}{cc}1&0\\0&e^{i \phi}\end{array}\right)  H^\dagger
\end{equation}
where $H$ is the Hadamard matrix of dimension two and $\phi \in [0, 2 \pi)$.
In this section we present theoretical probability of correct
discrimination between these measurements. To do that, we will present an auxiliary lemma.
\begin{lemma}\label{lemma:min-e-optimal}
	Let $U = H \diag(1, e^{i \phi}) H^\dagger$, $\phi \in [0, 2\pi)$ and	let
	$\Phi_U$ and $\Phi_\Id$ be two unitary channels. Then, the following equation holds
	\begin{equation}
	\min_{E \in \diaguni_2} \|\Phi_{UE} -
	\Phi_\Id\|_\diamond = \|\Phi_{U} -
	\Phi_\Id\|_\diamond,
	\end{equation}
\end{lemma}

\begin{proof} Recall that the distance between two unitary channels is given by
	$
	\| \Phi_U  - \Phi_{\1} \|_\diamond = 2 \sqrt{1-\nu^2},
	$
	where $\nu = \min_{x \in W(U^\dagger)} |x|  $ for any $U \in \mathcal{U}_d$.
	For $U = H
	\left(\begin{array}{cc}1&0\\0&e^{i \phi}\end{array}\right)  H^\dagger$ the readers briefly observe that  $\nu^2 = 1 - \frac{|1 - e^{-i \phi} |^2 }{4} = 1 - \frac{|1 - e^{i \phi} |^2 }{4}$. So,
	\begin{equation}
	\|  \Phi_U  - \Phi_{\1} \|_\diamond = | 1 - e^{i \phi} |.
	\end{equation}
	It implies that it is enough to prove  \begin{equation}
	\min_{E \in \diaguni_2} \|\Phi_{UE} -
	\Phi_\Id\|_\diamond  = | 1 - e^{i \phi} |.
	\end{equation}
	This condition is equivalent to show that for every $E \in \diaguni_2$
	\begin{equation}
	 \nu_{E} \le  \frac{|1 + e^{i \phi} | }{2},
	\end{equation}
	where $\nu_E = \min_{x \in W(U^\dagger E)} |x|. $

	The celebrated Hausdorf-T{\"o}plitz theorem~\cite{hausdorff, toeplitz} states that
	$W(A)$ of any matrix $A \in \mathcal{M}_d$ is a convex set, and therefore we have
	\begin{equation}
	W(A) = \{ \tr(A \rho): \rho \in \Omega_d\}.
	\end{equation}
	So, we can assume that
	\begin{equation}
	\min_{\ket{x} \in \mathbb{C}^2:   \proj{x} = 1} |\bra{x}U^\dagger\ket{x}| =
	\min_{\rho \in \Omega_2} |\tr(U^\dagger\rho)|.
	\end{equation}
	Then, we have
	\begin{equation}
	 \nu_{E}  =   \min_{\rho \in
		\Omega_2} \left| \tr \left( \rho U E \right) \right|.
	\end{equation}
	For that, our task is reduced to show that for every  $E \in \diaguni_2$ there exists $\rho \in \Omega_2$ such that
	\begin{equation}
 | \tr \left(\rho U E\right) | \le \frac{|1 + e^{i \phi} | }{2}.
	\end{equation}

	Let us define $E = \left(\begin{array}{cc}E_0&0\\0&E_1\end{array}\right)  $
	and take $\rho =
	\left(\begin{array}{cc}\frac{1}{2}&0\\0&\frac{1}{2}\end{array}\right) $.
	From spectral theorem, let us decompose $U$ as
	\begin{equation}
	U= \lambda_0 \ketbra{x_0}{x_0} + \lambda_1 \ketbra{x_1}{x_1},
	\end{equation}
	where  for eigenvalue $\lambda_0 = 1$, the corresponding
	eigenvector is
	of the form $\ket{x_0} = \left[\begin{array}{c}\frac{1}{\sqrt{2}}\\\frac{1}{\sqrt{2}}\end{array}\right]
	$,
	whereas for  $\lambda_1= e^{i \phi}$ we have $\ket{x_1} = \left[\begin{array}{c}\frac{1}{\sqrt{2}}\\-\frac{1}{\sqrt{2}}\end{array}\right]
	$.
	Then, for every $ E \in \diaguni_2 $ we have
	\begin{equation}
	\begin{split}
 | \tr (\rho U E) | & =   \frac{1}{2}  \left| \tr \left(
	H \diag(1, e^{i\phi}) H^\dagger E \right) \right| =
	\frac{1}{2} \left| \tr\left((   \proj{x_0} +e^{i \phi}\proj{x_1} ) E \right)
	\right| \\& =
	\frac{1}{2} \left|  \bra{x_0} E \ket{x_0} +  e^{i \phi}\bra{x_1} E \ket{x_1}
	\right| =
	\frac{1}{2} \left| \frac{E_0 + E_1}{2} + e^{i \phi } \frac{E_0+E_1}{2} \right|
	\\& =
	\frac{\left| 1+ e^{i \phi } \right|}{2} \left| \frac{E_0 + E_1}{2} \right| \le
\frac{|1 + e^{i \phi} | }{2},
	\end{split}
	\end{equation}
	which completes the proof.
\end{proof}

\begin{theorem}\label{th:probability}
	The optimal probability of correct discrimination between von Neumann
	measurements $\PP_U$ and $\PP_{\Id}$ for $U = H \diag(1, e^{i \phi}) H^\dagger$,
	where $\phi \in [0, 2\pi)$ is given by
	\begin{equation}
	p_{\text{succ}}(\PP_{U}, \PP_{\Id}) = \frac{1}{2} + \frac{|1 - e^{i \phi}  |}{4} .
	\end{equation}
\end{theorem}

\begin{proof}
	From Holevo-Helstrom theorem, we obtain
	\begin{equation}
	p_{\text{succ}}(\PP_{U}, \PP_{\Id}) = \frac{1}{2} + \frac{1}{4} \| \PP_{U} - \PP_{\Id} \|_\diamond.
	\end{equation}
	From~\cite[Theorem 1]{puchala2018strategies}, we have
	\begin{equation}
	\|\PP_U - \PP_\Id\|_\diamond = \min_{E \in \diaguni_d} \|\Phi_{UE} -
	\Phi_\Id\|_\diamond.
	\end{equation}
	From Lemma~\ref{lemma:min-e-optimal},  we know that for
	$U =  H \diag(1, e^{i \phi}) H^\dagger$,  it also holds that
	\begin{equation}
	\min_{E \in \diaguni_2} \|\Phi_{UE} -
	\Phi_\Id\|_\diamond = \|\Phi_{U} -
	\Phi_\Id\|_\diamond,
	\end{equation} which is exactly equal to
	\begin{equation}
	\|\Phi_{U} -
	\Phi_\Id\|_\diamond = 2\sqrt{1 - \nu^2} = |1-e^{i   \phi }|.
	\end{equation}
	It implies that
	\begin{equation}
	p_{\text{succ}}(\PP_{U}, \PP_{\Id}) = \frac{1}{2} + \frac{|1-e^{i \phi}|}{4},
	\end{equation} which completes the proof.
\end{proof}

\section{Optimal discrimination strategy for parameterized Fourier family} \label{app:optimal-strategy}

In this Appendix we create the optimal
theoretical strategy of  discrimination between $\PP_{U}$ and $\PP_{\Id}$.
To indicate the optimal strategy, we will present two propositions. The first one is concentrated around the discriminator as the optimal input state of discrimination strategy, whereas the second one describes the optimal final measurement.

\begin{proposition}\label{prop-discrim}
	Consider the problem of discrimination between von Neumann measurements $\PP_U$
	and $\PP_\1$, $U = H\diag(1, e^{i \phi}) H^\dagger $ and $\phi \in [0,
	2\pi)$.  The  discriminator has the form
	\begin{equation}
	\ket{\psi_{0}} = \frac{1}{\sqrt{2}} |\Id_2 \rangle \rangle.
	\end{equation}
\end{proposition}

\begin{proof}
	Let $U = H\diag(1, e^{i \phi}) H^\dagger, \phi \in [0,
	2\pi)$ be decomposed as
	\begin{equation}
	U= \lambda_0 \ketbra{x_0}{x_0} + \lambda_1 \ketbra{x_1}{x_1},
	\end{equation}
	where  for eigenvalue $\lambda_0 = 1$, the corresponding
	eigenvector is
	of the form $\ket{x_0} = \left[\begin{array}{c}\frac{1}{\sqrt{2}}\\\frac{1}{\sqrt{2}}\end{array}\right]
	$,
	whereas for  $\lambda_1 = e^{i \phi}$ we have $\ket{x_1} = \left[\begin{array}{c}\frac{1}{\sqrt{2}}\\-\frac{1}{\sqrt{2}}\end{array}\right]
	$.
	For Hermitian-preserving maps \cite{watrous} the diamond norm may be expressed as
	\begin{equation}
	\| \Phi  \|_\diamond =  \max_{\rho \in \Omega_d} \| \left( \Id \otimes \sqrt{\rho} \right) J(\Phi)  \left( \Id \otimes \sqrt{\rho} \right)  \|_1.  \end{equation}
	Hence, we obtain
	\begin{equation}
	\begin{split}
	\| \PP_{U} - \PP_{\Id}  \|_\diamond
	& =  \max_{\rho \in \Omega_2} \left\| \left( \Id \otimes \sqrt{\rho} \right)
	J(\PP_{U} - \PP_{\Id} )  \left( \Id \otimes \sqrt{\rho} \right)  \right\|_1
	\\
	& =  \max_{\rho \in \Omega_2} \left\| \left( \Id \otimes \sqrt{\rho} \right)
	\sum_{i=0}^{1} \proj{i} \otimes \left( \proj{u_i} - \proj{i} \right)^\top
\left( \Id \otimes \sqrt{\rho} \right) 	\right\|_1  \\
	& = \max_{\rho \in \Omega_2} \left\| \sum_{i=0}^{1} \proj{i} \otimes
	\sqrt{\rho}  \left( \proj{u_i} - \proj{i} \right)^\top \sqrt{\rho}  \right\|_1
	\\
	& = \max_{\rho \in \Omega_2} \sum_{i=0}^{1} \left\| \sqrt{\rho}  \left(
	\proj{u_i} - \proj{i} \right)^\top \sqrt{\rho}  \right\|_1.
	\end{split}
	\end{equation}
	One can prove that for all $\alpha, \beta \ge 0 $, and unit vectors $\ket{x},
	\ket{y}$ the following equation holds~\cite{watrous}
	\begin{equation}
	\| \alpha \proj{x} - \beta\proj{y} \|_1 = \sqrt{(\alpha + \beta)^2 - 4\alpha
		\beta |\braket{x}{y}|^2}.
	\end{equation}
	By taking $\ket{x} = \frac{\sqrt{\rho} \ket{\bar{u_i}}}{\| \sqrt{\rho}
		\ket{\bar{u_i}} \|}$ and $ \ket{y} = \frac{\sqrt{\rho} \ket{i}}{\|\sqrt{\rho}
		\ket{i} \|}$ we have
	\begin{equation}
	\| \PP_{U} - \PP_{\Id}  \|_\diamond  = \max_{\rho \in \Omega_2}
	\sum_{i=0}^{1} \sqrt{\left( \bra{\bar{u_i}} \rho \ket{\bar{u_i}} + \bra{i} \rho \ket{i
		}\right)^2 - 4 | \bra{\bar{u_i}} \rho \ket{i} |^2}.
	\end{equation}
	Let us take  $\rho_0 =   \frac{1}{2}
	\left(\begin{array}{cc}1&0\\0&1\end{array}\right)  $,   we obtain
	\begin{equation}
	\begin{split}
	||\mathcal{P}_U - \mathcal{P}_{\1}||_\diamond
	&\ge \sum_{i=0}^1
	\sqrt{\left(\bra{\bar{u_i}}\rho_0\ket{\bar{u_i}} + \bra{i} \rho_0 \ket{i} \right)^2 -
		4|\bra{i}\rho_0\ket{\bar{u_i}}|^2}  \\
	&= \sum_{i=0}^1  \sqrt{ 1 -  \left| \bra{i}  U \ket{i }\right|^2}
	\\
	&=\sum_{i=0}^1  \sqrt{1 -  \left| 1 \cdot \braket{i}{u_0}
		\braket{u_0}{i} + e^{i \phi} \cdot\braket{i}{u_1} \braket{u_1}{i}\right|^2} \\
	&= \sum_{i=0}^1
	\sqrt{1 -\left| \frac{1+ e^{i \phi}}{2}\right|^2 }
	= 2 \sqrt{1 -\left| \frac{1+e^{i \phi}}{2}\right|^2 } \\
	&= |1-e^{i \phi }|.
	\end{split}
	\end{equation}
	Due to Theorem \ref{th:probability} and the following equality
	\begin{equation}
	\begin{split}
 \| \left( \Id \otimes \sqrt{\rho} \right) J(\PP_{U} - \PP_{\Id} )  \left(
	\Id \otimes \sqrt{\rho} \right) \|_1 = \left\| ( (\PP_{U} - \PP_\Id) \otimes \Id) \left(  | \sqrt{\rho}^\top
	\rangle \rangle \langle \langle \sqrt{\rho}^\top | \right) \right\|_1,
	\end{split}
	\end{equation}
the discriminator $\ket{\psi_0}$
 is equal to  \begin{equation}
	\ket{\psi_{0}} =  | \sqrt{\rho_0}^\top
	\rangle \rangle  = \frac{1}{\sqrt{2} } |
	\Id_2 \rangle \rangle,
	\end{equation}
	which completes the proof.
\end{proof}

\begin{proposition}\label{prop:optimal-measurement}
	Consider the problem of discrimination between von Neumann measurements $\PP_U$
	and $\PP_\1$, $U = H\diag(1, e^{i \phi}) H^\dagger $ and $\phi \in [0,
	2\pi)$.
	The   controlled unitaries $V_0$ and $V_1$
	have the form
	\begin{equation}
	V_0 = \left(\begin{array}{cc}i \sin\left( \frac{\pi - \phi}{4} \right)&-i
	\cos\left( \frac{\pi - \phi}{4} \right)\\ \cos\left( \frac{\pi -
		\phi}{4}\right)& \sin\left( \frac{\pi - \phi}{4} \right)\end{array}\right),
	\end{equation}
	and
	\begin{equation}
	V_1 = \left(\begin{array}{cc}-i \cos\left(\frac{\pi - \phi}{4}\right) &i
	\sin\left( \frac{\pi - \phi}{4}\right)\\\sin\left( \frac{\pi - \phi}{4} \right)
	&  \cos\left( \frac{\pi - \phi}{4} \right) \end{array}\right).
	\end{equation}
\end{proposition}

\begin{proof}
	From Proposition~\ref{prop-discrim} we obtain the exact form of discriminator given by
	\begin{equation}
	\ket{\psi_0}  = \frac{1}{\sqrt{2}} |\Id_2
	\rangle \rangle.
	\end{equation}
	Repeating the procedure used to distinguish the von Neumann measurements in the Hadamard basis (see \ref{app:hadamard}), we use the Hahn-Jordan decomposition and then we create the projective operators into the positive and negative part of $X$ matrix.
	Hence, the explicit form of $V_0$ and $V_1$ is given as
	follows:
	\begin{equation}
	V_0 = \left(
	\begin{array}{cc}i \sin\left( \frac{\pi - \phi}{4} \right)&-i
	\cos\left( \frac{\pi - \phi}{4} \right)\\ \cos\left( \frac{\pi -
		\phi}{4}\right)& \sin\left( \frac{\pi - \phi}{4} \right)
	\end{array}
	\right),
	\end{equation}
and
	\begin{equation}
	V_1 = \left(\begin{array}{cc}-i \cos\left(\frac{\pi - \phi}{4}\right) &i
	\sin\left( \frac{\pi - \phi}{4}\right)\\\sin\left( \frac{\pi - \phi}{4}
	\right) &  \cos\left( \frac{\pi - \phi}{4} \right) \end{array}\right),
	\end{equation}
	where $\phi \in [0,2\pi)$.
\end{proof}

\section{Output YAML files} \label{app:example}
In this appendix we present examples of YAML's files obtained from synchronous ans asynchronous experiments. We will start at synchronous case.

\begin{lstlisting}[language=Python, caption=Defining experiment file]
type: discrimination-fourier
qubits:
	- target: 0
	  ancilla: 1
	- target: 1
	  ancilla: 2
angles:
	start: 0
	stop: 2 * pi
	num_steps: 3
gateset: ibmq
method: direct_sum
num_shots: 100
\end{lstlisting}

\begin{lstlisting}[language=Python, caption=Defining backend file]
name: ibmq_quito
asynchronous: false
provider:
	hub: ibm-q
	group: open
	project: main

\end{lstlisting}

\begin{lstlisting}[language=Python, caption=Results (synchronous)]
metadata:
	experiments:
		type: discrimination-fourier
		qubits:
		- {target: 0, ancilla: 1}
		- {target: 1, ancilla: 2}
		angles: {start: 0.0, stop: 6.283185307179586, num_steps: 3}
		gateset: ibmq
		method: direct_sum
		num_shots: 100
	backend_description:
		name: ibmq_quito
		asynchronous: false
		provider: {group: open, hub: ibm-q, project: main}
data:
- target: 0
  ancilla: 1
  phi: 0.0
  results_per_circuit:
  - name: id
	histogram: {'00': 28, '01': 26, '10': 21, '11': 25}
	mitigation_info:
		target: {prob_meas0_prep1: 0.052200000000000024, prob_meas1_prep0: 0.0172}
		ancilla: {prob_meas0_prep1: 0.05900000000000005, prob_meas1_prep0: 0.0202}
	mitigated_histogram: {'00': 0.2637212373658018, '01': 0.25865061319892463, '10': 0.2067279352110304, '11': 0.2709002142242433}
  - name: u
	histogram: {'00': 30, '01': 16, '10': 28, '11': 26}
	mitigation_info:
		target: {prob_meas0_prep1: 0.052200000000000024, prob_meas1_prep0: 0.0172}
		ancilla: {prob_meas0_prep1: 0.05900000000000005, prob_meas1_prep0: 0.0202}
	mitigated_histogram: {'00': 0.2857378991684036, '01': 0.14975297832942433, '10': 0.28142307224788693, '11': 0.2830860502542851}
- target: 0
  ancilla: 1
  phi: 3.141592653589793
  results_per_circuit:
  - name: id
	histogram: {'00': 4, '01': 5, '10': 45, '11': 46}
	mitigation_info:
		target: {prob_meas0_prep1: 0.052200000000000024, prob_meas1_prep0: 0.0172}
		ancilla: {prob_meas0_prep1: 0.05900000000000005, prob_meas1_prep0: 0.0202}
	mitigated_histogram: {'00': 0.011053610583159325, '01': 0.02261276648026373, '10': 0.4593955619936729, '11': 0.5069380609429042}
  - name: u
	histogram: {'00': 56, '01': 43, '10': 1}
	mitigation_info:
		target: {prob_meas0_prep1: 0.052200000000000024, prob_meas1_prep0: 0.0172}
		ancilla: {prob_meas0_prep1: 0.05900000000000005, prob_meas1_prep0: 0.0202}
	mitigated_histogram: {'00': 0.5573987337172156, '01': 0.44424718645642625, '10': -0.0016459201736417181}
- target: 0
  ancilla: 1
  phi: 6.283185307179586
  results_per_circuit:
  - name: id
	histogram: {'00': 36, '01': 18, '10': 25, '11': 21}
	mitigation_info:
		target: {prob_meas0_prep1: 0.052200000000000024, prob_meas1_prep0: 0.0172}
		ancilla: {prob_meas0_prep1: 0.05900000000000005, prob_meas1_prep0: 0.0202}
	mitigated_histogram: {'00': 0.3488190312089973, '01': 0.17355281935572894, '10': 0.2505792064871127, '11': 0.22704894294816103}
  - name: u
	histogram: {'00': 32, '01': 27, '10': 24, '11': 17}
	mitigation_info:
		target: {prob_meas0_prep1: 0.052200000000000024, prob_meas1_prep0: 0.0172}
		ancilla: {prob_meas0_prep1: 0.05900000000000005, prob_meas1_prep0: 0.0202}
	mitigated_histogram: {'00': 0.3025357275361897, '01': 0.27413673119534815, '10': 0.24313373302688793, '11': 0.18019380824157433}
- target: 1
  ancilla: 2
  phi: 0.0
  results_per_circuit:
  - name: id
	histogram: {'00': 27, '01': 20, '10': 24, '11': 29}
	mitigation_info:
		target: {prob_meas0_prep1: 0.05900000000000005, prob_meas1_prep0: 0.0202}
		ancilla: {prob_meas0_prep1: 0.07540000000000002, prob_meas1_prep0: 0.0528}
	mitigated_histogram: {'00': 0.2594378169217188, '01': 0.19318893233269735, '10': 0.23035366874292057, '11': 0.3170195820026633}
  - name: u
	histogram: {'00': 31, '01': 24, '10': 23, '11': 22}
	mitigation_info:
		target: {prob_meas0_prep1: 0.05900000000000005, prob_meas1_prep0: 0.0202}
		ancilla: {prob_meas0_prep1: 0.07540000000000002, prob_meas1_prep0: 0.0528}
	mitigated_histogram: {'00': 0.30056875246775644, '01': 0.2438221628798003, '10': 0.22180309809696985, '11': 0.23380598655547338}
- target: 1
  ancilla: 2
  phi: 3.141592653589793
  results_per_circuit:
  - name: id
	histogram: {'00': 5, '01': 4, '10': 50, '11': 41}
	mitigation_info:
		target: {prob_meas0_prep1: 0.05900000000000005, prob_meas1_prep0: 0.0202}
		ancilla: {prob_meas0_prep1: 0.07540000000000002, prob_meas1_prep0: 0.0528}
	mitigated_histogram: {'00': 0.009552870928837118, '01': 0.007194089383161034, '10': 0.5236791012692514, '11': 0.4595739384187503}
  - name: u
	histogram: {'00': 41, '01': 51, '10': 3, '11': 5}
	mitigation_info:
		target: {prob_meas0_prep1: 0.05900000000000005, prob_meas1_prep0: 0.0202}
		ancilla: {prob_meas0_prep1: 0.07540000000000002, prob_meas1_prep0: 0.0528}
	mitigated_histogram: {'00': 0.4073387714165384, '01': 0.5614614121117936, '10': 0.006431862814564833, '11': 0.024767953657102992}
- target: 1
  ancilla: 2
  phi: 6.283185307179586
  results_per_circuit:
  - name: id
	histogram: {'00': 30, '01': 28, '10': 23, '11': 19}
	mitigation_info:
		target: {prob_meas0_prep1: 0.05900000000000005, prob_meas1_prep0: 0.0202}
		ancilla: {prob_meas0_prep1: 0.07540000000000002, prob_meas1_prep0: 0.0528}
	mitigated_histogram: {'00': 0.2868459834940102, '01': 0.2919564941384742, '10': 0.22466574543735374, '11': 0.19653177693016174}
  - name: u
	histogram: {'00': 15, '01': 20, '10': 36, '11': 29}
	mitigation_info:
		target: {prob_meas0_prep1: 0.05900000000000005, prob_meas1_prep0: 0.0202}
		ancilla: {prob_meas0_prep1: 0.07540000000000002, prob_meas1_prep0: 0.0528}
	mitigated_histogram: {'00': 0.1187719606657805, '01': 0.1962085394489247, '10': 0.3710195249988589, '11': 0.31399997488643583}
\end{lstlisting}
For the same experiment file, we use the flag \texttt{asynchronous: true} to define asynchronous experiment.

\begin{lstlisting}[language=Python, caption=Backend file]
name: ibmq_quito
asynchronous: true
provider:
	hub: ibm-q
	group: open
	project: main

\end{lstlisting}
If the backend is asynchronous, the output will contain intermediate data
containing, amongst others, job ids correlated with the circuit they
correspond to.

\begin{lstlisting}[language=Python, caption=Resolved results]
metadata:
	experiments:
		type: discrimination-fourier
		qubits:
		- {target: 0, ancilla: 1}
		- {target: 1, ancilla: 2}
		angles: {start: 0.0, stop: 6.283185307179586, num_steps: 3}
		gateset: ibmq
		method: direct_sum
		num_shots: 100
	backend_description:
		name: ibmq_quito
		asynchronous: true
		provider: {group: open, hub: ibm-q, project: main}
data:
- job_id: 63e7f17a17b7ed49ca24e05b
  keys:
  - [0, 1, id, 0.0]
  - [0, 1, u, 0.0]
  - [0, 1, id, 3.141592653589793]
  - [0, 1, u, 3.141592653589793]
  - [0, 1, id, 6.283185307179586]
  - [0, 1, u, 6.283185307179586]
  - [1, 2, id, 0.0]
  - [1, 2, u, 0.0]
  - [1, 2, id, 3.141592653589793]
  - [1, 2, u, 3.141592653589793]
  - [1, 2, id, 6.283185307179586]
  - [1, 2, u, 6.283185307179586]

\end{lstlisting}
Finally, if the status of jobs is \texttt{DONE}, we resolve the measurements from the
submitted jobs obtaining the following file.
\begin{lstlisting}[language=Python, caption=Results (asynchronous)]
metadata:
  experiments:
	type: discrimination-fourier
	qubits:
	- target: 0
	  ancilla: 1
	- target: 1
	  ancilla: 2
	angles:
	  start: 0.0
	  stop: 6.283185307179586
	  num_steps: 3
	gateset: ibmq
	method: direct_sum
	num_shots: 100
  backend_description:
	name: ibmq_quito
	asynchronous: true
	provider:
		group: open
		hub: ibm-q
		project: main
data:
- target: 0
  ancilla: 1
  phi: 0.0
  results_per_circuit:
  - name: id
	histogram:
		'00': 27
		'01': 28
		'10': 18
		'11': 27
	mitigation_info:
	  target:
		prob_meas0_prep1: 0.052200000000000024
		prob_meas1_prep0: 0.0172
	  ancilla:
		prob_meas0_prep1: 0.05900000000000005
		prob_meas1_prep0: 0.0202
	mitigated_histogram:
		'00': 0.254196166145997
		'01': 0.2790358060520916
		'10': 0.1732699847244092
		'11': 0.29349804307750227
  - name: u
	histogram:
		'00': 29
		'01': 17
		'10': 30
		'11': 24
	mitigation_info:
	  target:
		prob_meas0_prep1: 0.052200000000000024
		prob_meas1_prep0: 0.0172
	  ancilla:
		prob_meas0_prep1: 0.05900000000000005
		prob_meas1_prep0: 0.0202
	mitigated_histogram:
		'00': 0.2733793468261183
		'01': 0.1621115306717096
		'10': 0.3045273800167787
		'11': 0.2599817424853933
- target: 0
  ancilla: 1
  phi: 3.141592653589793
  results_per_circuit:
  - name: id
	histogram:
		'00': 3
		'01': 5
		'10': 37
		'11': 55
	mitigation_info:
	  target:
		prob_meas0_prep1: 0.052200000000000024
		prob_meas1_prep0: 0.0172
	  ancilla:
		prob_meas0_prep1: 0.05900000000000005
		prob_meas1_prep0: 0.0202
	mitigated_histogram:
		'00': 0.006189545789708441
		'01': 0.016616709640352317
		'10': 0.3675478279476653
		'11': 0.6096459166222741
  - name: u
	histogram:
		'00': 56
		'01': 42
		'10': 2
	mitigation_info:
	  target:
		prob_meas0_prep1: 0.052200000000000024
		prob_meas1_prep0: 0.0172
	  ancilla:
		prob_meas0_prep1: 0.05900000000000005
		prob_meas1_prep0: 0.0202
	mitigated_histogram:
		'00': 0.55731929321128
		'01': 0.43367489257574243
		'10': 0.009005814212977551
- target: 0
  ancilla: 1
  phi: 6.283185307179586
  results_per_circuit:
  - name: id
	histogram:
		'00': 18
		'01': 28
		'10': 30
		'11': 24
	mitigation_info:
	  target:
		prob_meas0_prep1: 0.052200000000000024
		prob_meas1_prep0: 0.0172
	  ancilla:
		prob_meas0_prep1: 0.05900000000000005
		prob_meas1_prep0: 0.0202
	mitigated_histogram:
		'00': 0.15258295844557557
		'01': 0.2829079190522524
		'10': 0.3071204587046501
		'11': 0.25738866379752195
  - name: u
	histogram:
		'00': 32
		'01': 28
		'10': 23
		'11': 17
	mitigation_info:
	  target:
		prob_meas0_prep1: 0.052200000000000024
		prob_meas1_prep0: 0.0172
	  ancilla:
		prob_meas0_prep1: 0.05900000000000005
		prob_meas1_prep0: 0.0202
	mitigated_histogram:
		'00': 0.3026150836796529
		'01': 0.28491749668524724
		'10': 0.23230862145681827
		'11': 0.18015879817828173
- target: 1
  ancilla: 2
  phi: 0.0
  results_per_circuit:
  - name: id
	histogram:
		'00': 27
		'01': 16
		'10': 30
		'11': 27
	mitigation_info:
	  target:
		prob_meas0_prep1: 0.05900000000000005
		prob_meas1_prep0: 0.0202
	  ancilla:
		prob_meas0_prep1: 0.07540000000000002
		prob_meas1_prep0: 0.0528
	mitigated_histogram:
		'00': 0.256742095057232
		'01': 0.15000257115061383
		'10': 0.29821012040758116
		'11': 0.29504521338457296
  - name: u
	histogram:
		'00': 34
		'01': 22
		'10': 25
		'11': 19
	mitigation_info:
	 target:
		prob_meas0_prep1: 0.05900000000000005
		prob_meas1_prep0: 0.0202
	  ancilla:
		prob_meas0_prep1: 0.07540000000000002
		prob_meas1_prep0: 0.0528
	mitigated_histogram:
		'00': 0.3325088211394024
		'01': 0.22335261496979697
		'10': 0.2441636375921354
		'11': 0.19997492629866526
- target: 1
  ancilla: 2
  phi: 3.141592653589793
  results_per_circuit:
  - name: id
	histogram:
		'00': 3
		'01': 9
		'10': 51
		'11': 37
	mitigation_info:
	  target:
		prob_meas0_prep1: 0.05900000000000005
		prob_meas1_prep0: 0.0202
	  ancilla:
		prob_meas0_prep1: 0.07540000000000002
		prob_meas1_prep0: 0.0528
	mitigated_histogram:
		'00': -0.016627023111853642
		'01': 0.06778554570877951
		'10': 0.53899887367658
		'11': 0.40984260372649417
  - name: u
	histogram:
		'00': 43
		'01': 45
		'10': 7
		'11': 5
	mitigation_info:
	  target:
		prob_meas0_prep1: 0.05900000000000005
		prob_meas1_prep0: 0.0202
	  ancilla:
		prob_meas0_prep1: 0.07540000000000002
		prob_meas1_prep0: 0.0528
	mitigated_histogram:
		'00': 0.42955729968594086
		'01': 0.49336080079582095
		'10': 0.04937406434533623
		'11': 0.02770783517290191
- target: 1
  ancilla: 2
  phi: 6.283185307179586
  results_per_circuit:
  - name: id
	histogram:
		'00': 22
		'01': 19
		'10': 35
		'11': 24
	mitigation_info:
	  target:
		prob_meas0_prep1: 0.05900000000000005
		prob_meas1_prep0: 0.0202
	  ancilla:
		prob_meas0_prep1: 0.07540000000000002
		prob_meas1_prep0: 0.0528
	mitigated_histogram:
		'00': 0.19592641048040849
		'01': 0.18787721420415215
		'10': 0.3590258049844047
		'11': 0.25717057033103463
  - name: u
		histogram:
		'00': 27
		'01': 24
		'10': 25
		'11': 24
	mitigation_info:
	  target:
		prob_meas0_prep1: 0.05900000000000005
		prob_meas1_prep0: 0.0202
	  ancilla:
		prob_meas0_prep1: 0.07540000000000002
		prob_meas1_prep0: 0.0528
	mitigated_histogram:
		'00': 0.25555866817587225
		'01': 0.2429501641251142
		'10': 0.24509293912212946
		'11': 0.2563982285768841
\end{lstlisting}

\end{document}